\documentclass[12pt,preprint]{AwadSh_aastex}

\usepackage{graphicx}	
\usepackage{amsmath}	
\usepackage{amssymb}	
\usepackage{longtable}
\usepackage{rotate}
\usepackage{lscape}
\usepackage{latexsym,verbatim}
\slugcomment{Accepted in Ap\&SS - printed on \today}
\shorttitle{Deuteration in NGC 2264 CMM3}
\shortauthors{Awad \& Shalabiea}
\begin{document}
\title{Deuterium Chemistry in the Young Massive Protostellar Core NGC 2264 CMM3}
\author{Z. Awad \and O. M. Shalabiea}
\affil{Astronomy, Space Science and Meteorology Department, Faculty of Science, Cairo University, Giza, Egypt} 
\email{zma@sci.cu.edu.eg; shalabiea@sci.cu.edu.eg}
\begin{abstract}

In this work we present the first attempt of modelling the deuterium chemistry in the massive young protostellar core NGC 2264 CMM3. We investigated the sensitivity of this chemistry to the physical conditions in its surrounding environment. The results showed that deuteration, in the protostellar gas, is affected by variations in the core density, the amount of gas depletion onto grain surfaces, the CR ionisation rate, but it is insensitive to variations in the H$_2$ ortho-to-para ratio. 

Our results, also, showed that deuteration is often enhanced in less-dense, partially depleted ($<$ 85\%), or cores that are exerted to high CR ionisation rates ($\ge$ 6.5 $\times$ 10$^{-17}$ s$^{-1}$ ). 
However, in NGC 2264 CMM3, decreasing the amount of gas depleted onto grains and enhancing the CR ionisation rate are often overestimating the observed values in the core. The best fit time to observations occurs around (1 - 5) $\times$ 10$^4$ yrs for core densities in the range (1 - 5) $\times$ 10$^6$ cm$^{-3}$ with CR ionisation rate between (1.7 - 6.5)$\times$ 10$^{-17}$ s$^{-1}$. 
These values are in agreement with the results of the most recent theoretical chemical model of CMM3, and the time range of best fit is, also, in-line with the estimated age of young protostellar objects.

We conclude that deuterium chemistry in protostellar cores is: 
(i) sensitive to variations in the physical conditions in its environment, 
(ii) insensitive to changes in the H$_2$ ortho-to-para ratio. 
We also conclude that the core NGC 2264 CMM3 is in its early stages of chemical evolution with an estimated age of (1 - 5) $\times$ 10$^4$ yrs.

\end{abstract}
\keywords{
Astrochemistry - Stars: massive, protostars, formation - ISM: abundances, molecules
}
\section{Introduction}
\label{intro}
Although the standard interstellar atomic deuterium-to-hydrogen ratio (D/H ratio) is low ($\sim$ 10$^{-5}$; \citealt{oli03}), observations revealed that this ratio is increased in several astrophysical regions by few orders of magnitudes (e.g. \citealt{tin20, cec01, loi01, crap05, cec07, sak07, cout13, vas17}). This phenomena is known as super-deuteration (D/H $>$ 10\%). It is obtained for some species such as H$_2$CO and CH$_3$OH in low-mass protostars (e.g. \citealt{cec07} and references therein). To-date, more than 30 deuterated species have been observed in different astrophysical regions with abundances comparable, sometimes, to their normal hydrogen counterparts (also known as main isotope). These species are ranged from singly (e.g. HD, H$_2$D$^+$, DCO$^+$, CH$_2$DOH, HDO, NH$_2$D, DCOOCH$_3$) to multiply\footnote{{\bf Singly (or mono-) deuterated:} are species in which only one hydrogen atom is replaced by a deuterium while {\bf multiply deuterated:} are species in which two or more hydrogen atoms are replaced.} deuterated species (e.g. D$_2$H$^+$, D$_2$CO, CHD$_2$OH, CD$_3$OH, NHD$_2$, ND$_3$). Details on observations of deuterated species in different astrophysical regions can be found in reviews such as \citealt{her09, cas012, tiel13}. Also, a list of most of the observed deuterated species in cores in massive and low-mass star forming regions is given in Table 1 in \citet{awad14} while Table 19 in \citet{albrt013} summarises the molecular D/H ratios in different interstellar environments. 

In the last few decades, numerous models were published to study the deuterium chemistry in the interstellar medium (ISM). Early attempts of modelling deuterium chemistry relayed on either pure gas-phase \citep{wats80} or surface chemistry \citep{tiel83}. \citet{bro89b,bro89a} introduced the first gas-grain chemical model to study deuterium chemistry in dense regions. After that, \citet{mil89} published a model of a particular importance because its results highlighted the temperature dependence of the main source of deuterium fractionation\footnote{{\bf Fractionation} is usually defined as the ratio between the molecule XH and its deuterated counterpart XD. However, sometimes, it refers to the amount of deuterated species formed from a given pathway. The latter definition is what we use in this work.}. 
These findings showed that in cold regions (T $\sim$ 10 K) H$_2$D$^+$ and its daughter ions DCO$^+$ and H$_2$DO$^+$ led to high fractionation while in warmer regions (10 $<$ T(K) $<$ 70) sources of high deuteration are CH$_2$D$^+$, C$_2$HD$^+$, and associated species. Years after, \citet{rod96} gave the first recipe to extend a given gas-phase chemical network to include deuterated species by assuming that hydrogenated and their deuterated counterparts react with the same rate coefficients. More recently, \citet{aik12} published another method for extending chemical networks in which they rely on statistical probabilities in determining the rate of a deuterated reaction from its original rate of reaction. Both recipes have been used in several studies (e.g. \citealt{rob20a, rob20b, rob03, rob04, flo06, sip10, albrt013, cout13, cout14, awad14, hin14, fur16, maj17}). 
The effect of including the spin state (ortho, para, \& meta) of species on the deuterium fractionation in cold environments was also studied and it is found to influence the fractionation (e.g. \citealt{gerl02, wal04, flo06, sip10, hin14}). 

The massive cluster-forming region NGC 2264 was a target of several recent studied to investigate its dynamics and morphology (e.g. \citealt{per06, per07, mau09, sar11}) and chemistry \citep{wat15}. The analysis of the spectral survey by \citet{wat15} identified about 36 species in the young massive protostellar core NGC 2264 CMM3 among which seven deuterated species were detected. These species are DCO$^+$, N$_2$D$^+$, DCN, DNC, CCD, HDCO, and DC$_3$N. All of these molecules have been detected, separately, in dark clouds \citep{lor85, vant09}, prestellar cores and protoplanetary disks \citep{mie12, vand03, ob15, hua15, vant09}, and in star-forming regions \citep{vand95, sak09, berg11}. In CMM3, neither singly deuterated methanol nor multiply deuterated species were reported \citep{wat15}. The authors contributed this non-detection to the limitation and insufficient sensitivity of their millimetre line survey ($\sim$ 30 mK at the corresponding frequency of CH$_2$DOH; \citealt{wat15}). 

The most recent observations by \citet{wat15} motivated \citet{awad17} to introduce the first gas-grain astrochemical model of CMM3 that looked into the chemical evolution of a selected set of these species that do not include deuterium. The identification of few deuterated species in CMM3 motivated us to modify this model and update its chemical network to include all possible deuterated species in order to understand their chemical evolution of in CMM3. Therefore, the main aim of this work is to investigate the sensitivity of deuterium chemistry in protostellar cores, specifically CMM3, to variations in the physical parameters namely; the density, the depletion onto grain surfaces, and the CR ionisation rate. Moreover, we examined the effect on changing the H$_2$ ortho-to-para ratio (OPR) on the deuterium chemistry in the protostellar phase.

This paper is organised as follows: we describe the used chemical model and the initial conditions in \S \ref{mod}, the results are discussed in detail in \S \ref{res}, and finally our concluding remarks are summarised in \S \ref{conc}. 

\section{Chemical Models and Initial Conditions}
\label{mod}
The chemical model used in the present work is based on the previously published model by \citet{awad17} for the protostellar core NGC 2264 CMM3 , but with a modified chemical network that includes all the possible forms of deuterated species of a given normal isotope. 
Briefly, the model used is a single point, time-dependent, gas-grain astrochemical model. CMM3 is a protostellar object hence the formation of molecules has been started before its formation. For this reason, we start following the chemistry of CMM3 after running a pre-phase under dark cloud conditions to obtain the initial input chemistry for the protostellar phase. In this pre-phase, we follow the chemistry of low-density gas ($\sim$ 400 cm$^{-3}$), initially atomic, undergoes a free fall collapse, following \citet{raw92}, at 10 K to reach a final density 5.4 $\times$ 10$^6$ cm$^{-3}$, which is observed for CMM3, in 10$^6$ yrs. The chemistry takes place both in gas-phase and on grain surfaces with the standard values of radiation field (1 Draine) and CR ionisation rate ($\zeta_{{\text {ISM}}}$ = 1.3 $\times$ 10$^{-17}$ s$^{-1}$). Following \citet{raw92} and \citet{snow06}, the visual extinction, A$_v$, in this phase varies as function of the gas density as 
$$ A_v ~(\text {mag})= \frac {\text {gas density (cm$^{-3}$)} \times ~\text {core size (cm)}} {N(H) ~\text {(cm$^{-2}$)}}$$ where N(H) is the standard column density of hydrogen at A$_v$ of 1 magnitude; 1.6 $\times$ 10$^{21}$ cm$^{-2}$. The results of the final time step of this pre-phase is considered as the initial input for the chemistry occurs in the protostellar phase. By the end of the prestellar phase, most of the species were frozen onto grain surfaces. 

The chemistry, in the protostellar phase, is then followed in a steady core of uniform density of 5.4 $\times$ 10$^6$ cm$^{-3}$ \citep{per06}, corresponds to a point of A$_v$ = 314 mag calculated self-consistency in the code. This value is in agreement with the observed range of visual extinctions in the region NGC 2264 \citep{war20}. The core has a temperature of 15 K 
and is irradiated by the standard interstellar radiation field of 1 Draine, and the cosmic ray ionisation reactions occur at a rate of $\zeta_{{\text {CMM3}}}$ = 1.67 $\times$ 10$^{-17}$ s$^{-1}$ \citep{awad17}. The physical conditions of CMM3 and the initial chemical elemental abundances are listed in Table \ref{tab:initial}. 

The chemistry takes place both in gas-phase and on grain surfaces. The gas-phase chemical network is adapted from the latest release of KIDA astrochemical database\footnote{KIDA: KInetic Database for Astrochemistry (http://kida.obs.u-bordeaux1.fr). We used the rate file kida.uva.2014} for dense media \citep{wak15}. The reactions of deuterated species are generated from their normal counterparts following the methodology described in details in \citet{aik12} and \citet{fur13}. In addition, we have included all the multiply deuterated form of H$_3^+$ for their important role in deuteration \citep{rob03}. We have also added the nuclear spin states of H$_2$, H$_3^+$, and their isotopologues (e.g. \citealt{flo06, sip10, hin14, cout14}). We considered the statistical ortho-to-para ratio of H$_2$ of 3 \citep{cout14, hin14}. 
 The surface chemical network is mainly simple hydrogenation and deuteration of the depleted species, in addition to the other surface reactions described in details in \citet{awad17}. 
Mantle species are returned to the gas-phase by non-thermal desorption mechanisms; H$_2$ formation, CR-photodissociation, and CR induced photodesorption \citep{robj07}. Thermal evaporation is not efficient in such low temperature (15 K). 

In addition to the reference model ({\it hereafter} the RM), we performed six other models (M1 to M6) to investigate how the core deuterium chemistry is affected by its surrounding physical conditions. 
Table \ref{tab:grid} lists the grid of these models; giving the values of the three physical parameters we studied compared to the reference model (in the first raw). In addition to those six models, we performed few models by which we investigated the effect of varying the H$_2$ OPR from the statistical ratio of 3 down to 0.003, following \citet{pag11} on the deuterium chemistry of the protostar. The chemical network in all models includes 191 species linked in 8180 gas-phase as well as surface reactions. The fractional abundances (with respect to the total hydrogen in all forms) are computed in the code by the 
rate equation method (e.g. \citealt{cup17}). In this method, the chemical network of each molecule is converted into a single ordinary differential equation (ODE) that describes the time evolution of this molecule. This equation, for a species X, has the general form $$ \frac {d[X]}{dt} = \sum{F(X) + D(X)} $$ where [X] is the concentration of the species X and F(X) and D(X) refers to the formation and destruction pathways of X, respectively. In order to solve the whole set of these ODEs and obtain the abundances for all the species involved in the network, we use the FORTRAN package LSODE with a double precision version (DLSODE) in which the ODEs are solved and directly integrated for all the species at each time step. 
The rate constants of the reactions are calculated automatically in the code using the UMIST formula \citep{mil97,mce013}.

\section{Results and Discussion}
\label{res} 
In this section we represent and discuss our model results for the protostellar phase. We start with reviewing, in details, the results of the RM (\S \ref{RM}), after that we discuss the effects of varying the physical conditions of the protostellar core on the chemical evolution of the deuterium chemistry (\S \ref{phys}). 
\subsection{The Reference Model}
\label{RM} 
\begin{figure*}
\begin{center} 
\includegraphics[trim= 1.4cm 0.8cm 1.5cm 1.75cm,clip=true,width=14cm]{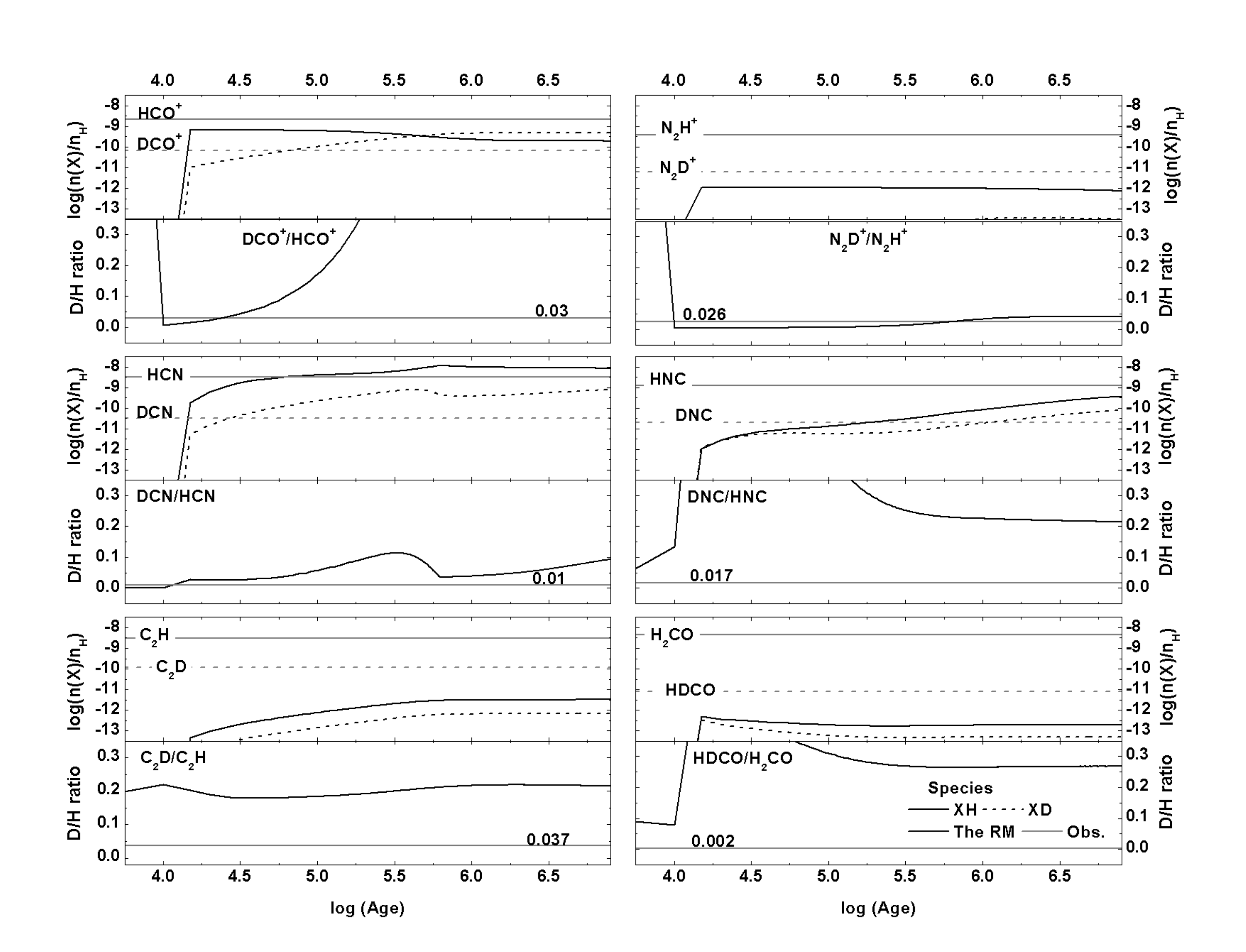}  
\includegraphics[trim= 1.8cm 1.1cm 1.5cm 1.6cm,clip=true,width=10cm]{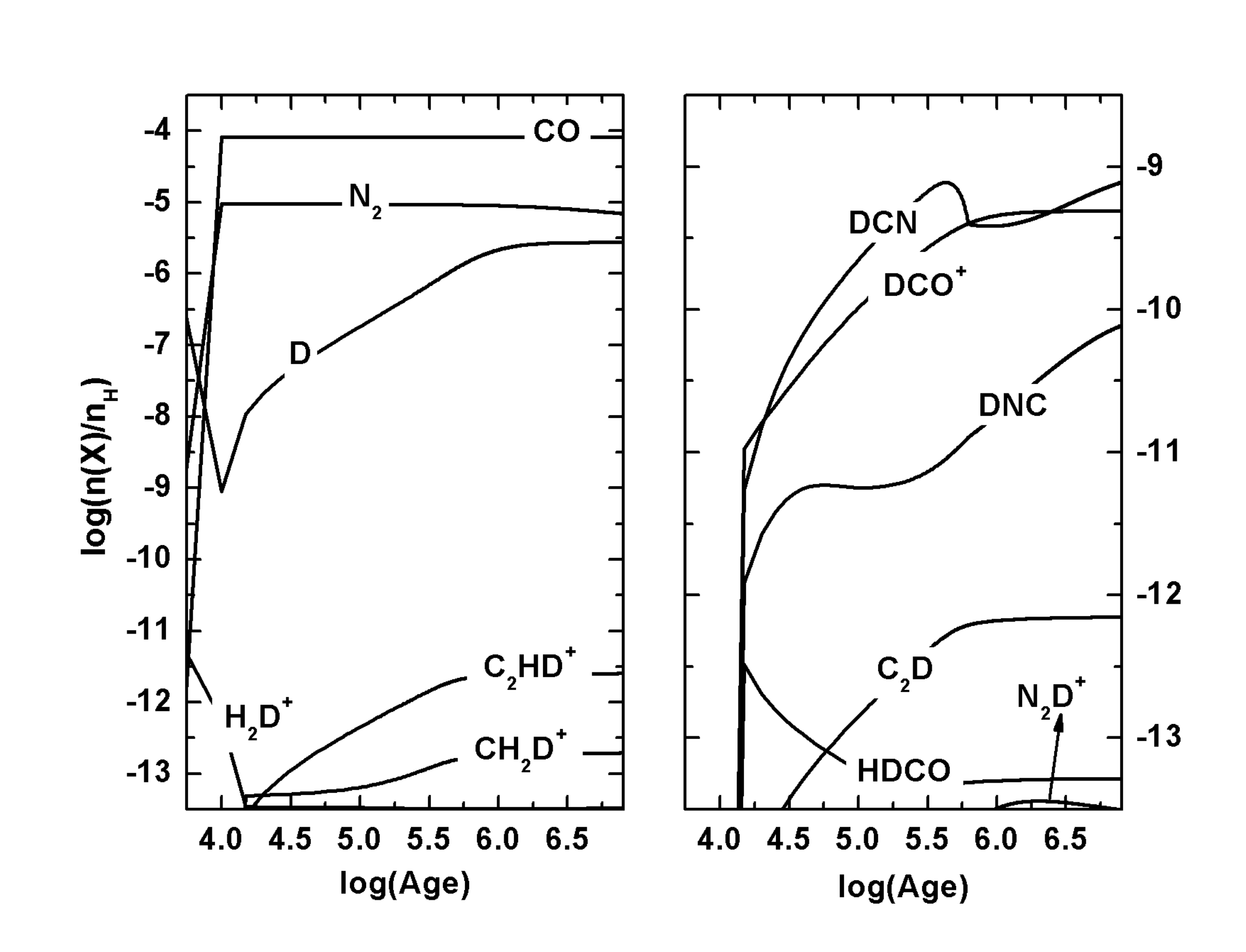}  
\caption {Top Panel: abundances and D/H ratios as a function of time as calculated in the reference model (RM) of CMM3. Solid lines represents normal isotopes (XH) while dashed lines are their deuterated counterparts (XD). Black lines represents the model calculations and grey ones are observations as taken from \citet{wat15}. The numbers typed in the ratio panels are the observed value. Bottom Panel: the time evolution of deuterated species with that of key deuterium chemistry drivers (see labels).}
\label{fig:rm}
\end{center}
\end{figure*}

Our reference model (RM) ran under the physical conditions of CMM3 listed in Table \ref{tab:initial} after a pre-phase of a dark cloud to get the initial molecular abundances of the protostellar phase. Fig. \ref{fig:rm} represents, in the top panel, the time evolution of the molecular abundances of our model calculations, in the protostellar phase, for a selected set of deuterated species (DCO$^+$, N$_2$D$^+$, DCN, DNC, C$_2$D, and HDCO) observed in CMM3 by \citet{wat15} and their normal isotopes (XH). For each molecule, we show the time evolution of the D/H ratio with comparison with the observed value. The bottom panel of Fig. \ref{fig:rm} is the time evolution of the set of selected deuterated molecules with some key species that derive the deuterium chemistry (D-atoms, CO, N$_2$, H$_2$D$^+$, CH$_2$D$^+$, and C$_2$HD$^+$). Note that the obtained increase in the abundance of some of these species is mainly due to the high CR ionisation rate in the protostellar phase (1.3 times higher than the dark cloud phase). The influence of this higher value is twofold; destruction of some gaseous species in the medium leading to an enhancement in the abundances of others such as the case of DCO$^+$ that is enhanced by the destruction of H$_2$D$^+$ and CH$_4$D$^+$ by CO, and releasing a small fraction of mantle species to desorb from grain surfaces to enrich the gas-phase such as the case of CO.

From Fig. \ref{fig:rm} one may notice similarities between the evolutionary trends of hydrogenated (XH) and their deuterated counterparts (XD). This may indicate either common chemical pathway (e.g. HCO$^+$ + D $\rightarrow$ DCO$^+$) or parent molecule such as (C$_2$HD$^+$ + NH$_3$ $\rightarrow$ NH$_4^+$ + C$_2$D or NH$_3$D$^+$ + C$_2$H). We also notice that the abundances of the species was reproducible up to a factor of 5 of the observations in particular for some XH species such as HCO$^+$, HCN, and HNC and their deuterated counterparts, but it was not successful in forming N$_2$D$^+$, C$_2$D, and HDCO and their XH isotopes. The evolutionary trends of the deuterium chemistry drivers (Fig. \ref{fig:rm}, bottom panel) and those of the studied molecules with the chemical analysis of the species may explain these results. 

The chemical analysis of the RM revealed that, at any given time, the formation of N$_2$D$^+$ proceeds via the reaction of N$_2$ and H$_2$D$^+$ and it is destroyed by CO molecules. Since CO is very abundant in the gas ($\sim$ 10$^{-4}$ with respect to the total H in the gas) one should expect a low abundance of H$_2$D$^+$ (e.g. \citealt{pag11}; $<$ 10$^{-11}$ in our RM) which could cause the less production of N$_2$D$^+$. Moreover, the high abundance of CO leads to a severe destruction of the ion with which the formation is not capable of maintaining the ion in the gas. Therefore, we barely (in the model) see the ion. The abundances of both C$_2$D and HDCO are comparable to those of C$_2$HD$^+$ and CH$_2$D$^+$, respectively. The chemical analysis showed that C$_2$D and HDCO are daughters of C$_2$HD$^+$ and CH$_2$D$^+$, respectively, and hence their low abundance is due to the shortage in the amount of their parents $\sim$ 5 $\times$ 10$^{-12}$). More details on the analysis of C$_2$D and HDCO is discussed in \S \ref{lim}.

DCO$^+$ and DCN best match observations at times (2 - 6) $\times$ 10$^4$ yrs while the time of best fit of DNC is slightly later ($\sim$ 4 $\times$ 10$^4$ yrs) due to the lack of its parent molecules (CHD and CD$_2$) during earlier times. The time of best fit is within the expected age of CMM3 \citep{awad17}. 

Although the model was not very successful in reproducing the fractional abundances of all the selected set of species, it was successful in reproducing most of the D/H ratios. The D/H ratio of HCO$^+$ is higher than 1 during times $<$ 10$^4$ and it saturates at 2 after 5 $\times$ 10$^5$ yrs; when the abundance of DCO$^+$ become twice that of HCO$^+$. The ratio of N$_2$H$^+$ also starts with values higher than unity then it declines to saturate at value comparable to observations ($\sim$ 0.03). D/H ratios $>$ 1 are omitted because they do not have real physical meaning since most of them are obtained when the species, both or one of them, are below the detection limit ($x$(X) $<$ 10$^{-13}$). When the species are detectable, we found that the ratios are in good agreement with observations during early stages, apart from those of C$_2$D and HDCO which are higher than observations due to their underestimated abundances. 

The observed D/H ratio of H$_2$CO in CMM3 is $\sim$ 10 times less than that obtained for other species. From our study, we expect that the reason for this is the low temperature of the gas (15K) which is not enough to evaporate the species from the dust grains since formaldehyde is believed to be effectively formed on grains (e.g. \citealt{tiel82, tiel83, wat03}) beside its formation in the gas. Another plausible reason for this low ratio is the chemical youth of CMM3 which does not allow time for more complex organics to form given the low abundance of their deuterium initiator (C$_2$HD$^+$ and CH$_2$D$^+$). The last point is related to the sensitivity of the detector which was the argument of \citet{wat15} for not detecting multiply deuterated species as well as singly deuterated methanol. It is possible that more sensitive detectors will be able to detect more lines of H$_2$CO and its deuterium counterparts, and hence enhance their abundances and then improve the ratio. 

A comparison between our computed abundances and the molecular D/H ratios and observations as taken from \citet{wat15} is given in Table \ref{tab:comp}.

\subsection{The Effect of Changing the Physical Parameters}
\label{phys} 
We investigated the impact of changing the following parameters: (1) the core density, (2) the amount of depletion onto grain surfaces, (3) the CR ionisation rate, and (4) H$_2$ OPR, on the fractional abundances of DCO$^+$, N$_2$D$^+$, DCN, DNC, C$_2$D, \& HDCO that had been observed in CMM3 by \citet{wat15}. 

Generally, we find that deuterium-bearing species are affected by changes in the parameter space we explored in this study. Most of the molecular abundances are enhanced up to 400 times (e.g. models M1, M3, M4), in particular during early stages of the evolution (t $\le$ 3 $\times$ 10$^4$ yrs) when compared to the results of the RM. Few evolutionary trends show similarities to those of the RM (e.g. models M2, M5, and M6) which may indicate similar pathways, but different rates. Detailed discussion of our findings is given in the following subsections. In our discussion we give the full chemical analysis of DCO$^+$ and DCN as an example of the ions and simple molecules, respectively, in the core. 

\subsubsection{The Core Density}
\label{den} 
\begin{figure*}
\begin{center} 
\includegraphics[trim= 1.5cm 0cm 1.0cm 0cm,clip=true,width=18cm]{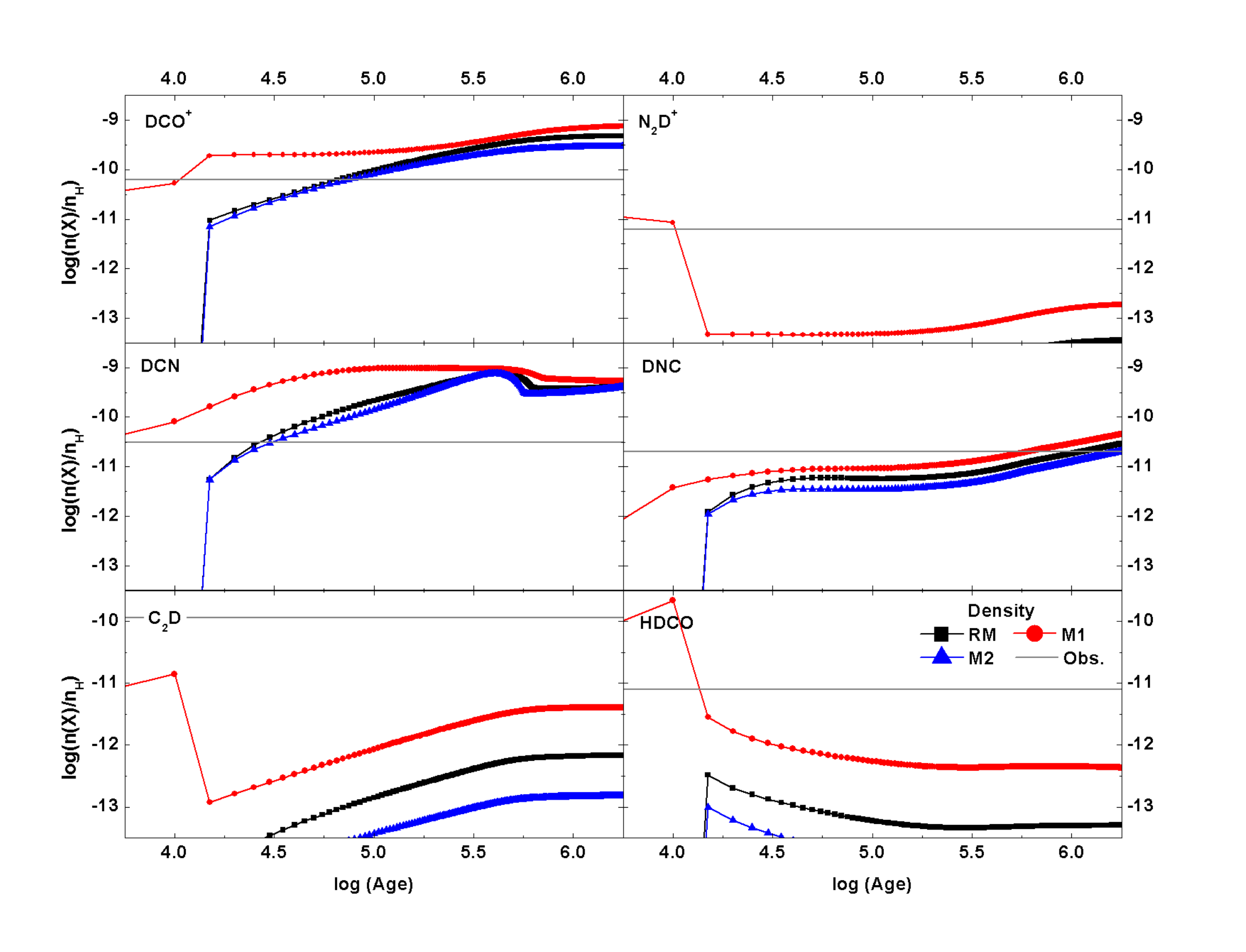}  
\caption {Abundances as a function of time in cores with different densities; 5.4$\times$10$^6$ cm$^{-3}$ (RM, black squares), 1.0$\times$10$^6$ cm$^{-3}$ (M1, red circles) and 2.0$\times$10$^7$ cm$^{-3}$ (M2, blue triangles). Observed values, as taken from \citet{wat15}, are represented by solid grey line.}
\label{fig:1}
\end{center}
\end{figure*}

In the context we are discussing the influence of changing the density of the core on the time evolution of the species understudy. Beside the reference model (RM) in which the density is 5.4 $\times$ 10$^6$ cm$^{-3}$ \citep{per06}, we performed two models; model M1 in which the density is 5 times less than that of the RM (n$_{\text H}$(M1) = 1.0 $\times$ 10$^6$ cm$^{-3}$), and model M2 where the density (n$_{\text H}$(M2) = 2.0 $\times$ 10$^7$ cm$^{-3}$) is 5 times higher than the density of the RM. The results of these models, M1 and M2, in comparison with those of the RM are illustrated in Fig. \ref{fig:1}. In this figure, black squares represent the RM while red circles and blue triangles denote models M1 and M2, respectively.

\begin{figure*}
\begin{center} 
\includegraphics[width=16cm]{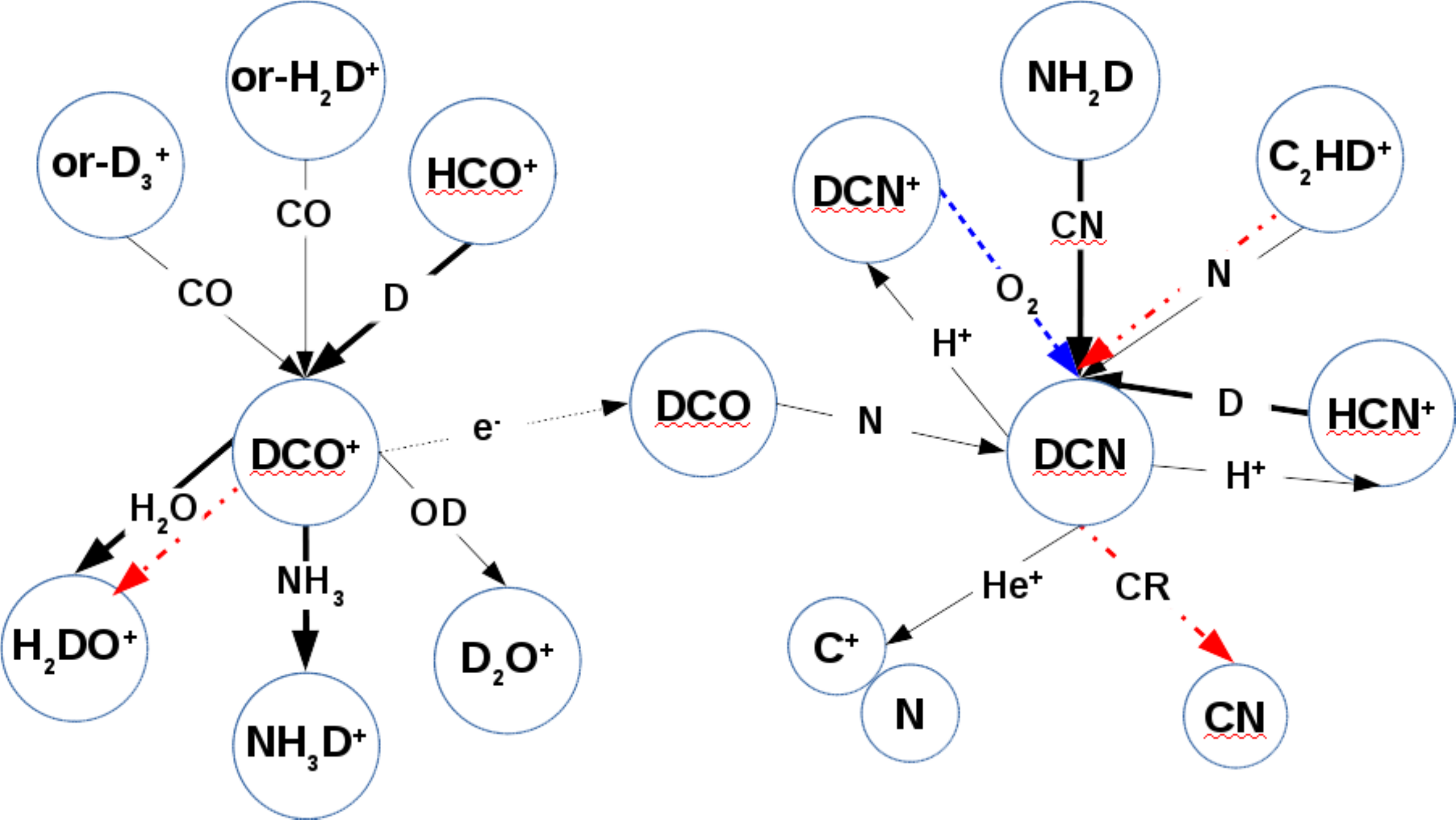}  
\caption {The effective chemical pathways for DCO$^+$ and DCN in CMM3 as given by our model chemical analysis in the RM (black arrows), low density (red dash-dot arrows) and low depletion (blue dashed line); see text in \S\ref{res}. {\bf The prefix `or-' stands for ortho.}} 
\label{fig:1.1}
\end{center}
\end{figure*}

In general, our calculations showed that reducing the core density (model M1) enhances the abundances of most of the species up to few orders of magnitude while increasing the core density (model M2) leads to an insignificant (often less than a factor of 5) decrease in the abundances of the species, see Fig. \ref{fig:1}. This increase in the abundances in model M1 is more pronounced during early evolutionary stages (t $\le$ 3 $\times$ 10$^4$ yrs) after which the abundances converge to those of the RM at later times. The overall chemical analysis of the species in model M1 showed that their rates of formation are few orders of magnitude higher than those in the RM, in particular at t $<$ 10$^4$ yrs, while their rates of destruction are within a factor of 5 of that calculated for the RM. 
Similarities between the evolutionary curves of model M2 and the RM indicate similarities in the chemical pathways, and differences in the rates of formation of species which is found to be a factor of 3 - 5 lower that those of the RM. Differences in the rates of formation and destruction between the three models could be attributed to differences in the abundances of the species parent molecules and their destroyers.
In the following we discuss the detailed chemical analysis of DCO$^+$ and DCN as examples to understand how the chemistry is affected by variations in the core density. In Fig. \ref{fig:1.1} we sketch the chemical network of both DCO$^+$ and DCN as revealed from the RM. Additional pathways with different arrow line styles indicates other routes of formation and/or destruction of the species under various physical conditions in the six models we performed.

In the RM, the formation of DCO$^+$ is dominated by the ion-molecule reactions of H$_3^+$ isotopes with CO at t $<$ 3 $\times$ 10$^4$ yrs, and the H-D substitution reaction `HCO$^+$ + D' at later times. 
The destruction of DCO$^+$ always occur by a competition between water (H$_2$O) and ammonia (NH$_3$). The same network is found to dominate the chemistry in models M1 and M2 with different times. The H-D substitution reaction is efficient since early stages (t $<$ 10$^4$ yrs) giving an increase to the productivity of DCO$^+$ during this time interval compared to the RM and hence enhance the formation rate. At times $\ge$ 10$^5$ yrs, the chemistry in the RM and M1 become comparable which explains the conversion of the results at late times. Moreover, the abundances of H$_2$O and NH$_3$ in the RM is higher which may also explain the lower abundance of the molecule in the RM. In denser gas (model M2), the abundances of the destroyers are slightly higher than those in the RM that in turn led to higher rates of destruction of DCO$^+$ (factor of 3 - 5 of the RM) while the comparable rates of formation could be attributed to the comparable abundances of the parent molecules in models M2 and the RM.

The same scenario is applicable for the simple molecule DCN in which its formation in the RM occurs via `HCN$^+$ + D' and `CN + NH$_2$D' and it is mainly destroyed by H$^+$. In model M1 the formation occur by these two routes in addition to `DCN$^+$ + O$_2$' at t $<$ 10$^4$ yrs; see Fig. \ref{fig:1.1}. This may increase the yield of the molecule $\sim$ 200 times than the RM during such early stages. The fact that the abundance of the ion H$^+$, the main cause of destroying DCN, is negligible in model M1 ($\sim$ 10$^{-11}$) compared to the RM ($\sim$ 10$^{-8}$) during t $<$ 10$^4$ yrs can be another reason for the high abundance of DCN during early times.

Observations of the massive protostellar core CMM3 \citep{sar11, wat15} suggested that the core is in its early evolutionary stages. Our results (Fig. \ref{fig:1}) showed generally that most of the abundances of the species are in good agreement with observations in the RM at times around (1 - 5) $\times$ 10$^4$ yr, except DNC which is in-line with observations at times $\ge$ 10$^5$ yrs. These times are in agreement with the ages of protostellar objects estimated by \citet{and93, and20}. 

\subsubsection{Depletion of the Gas}
\label{dep} 
\begin{figure*}
\begin{center} 
\includegraphics[trim= 1.5cm 0cm 1.0cm 0cm,clip=true,width=18cm]{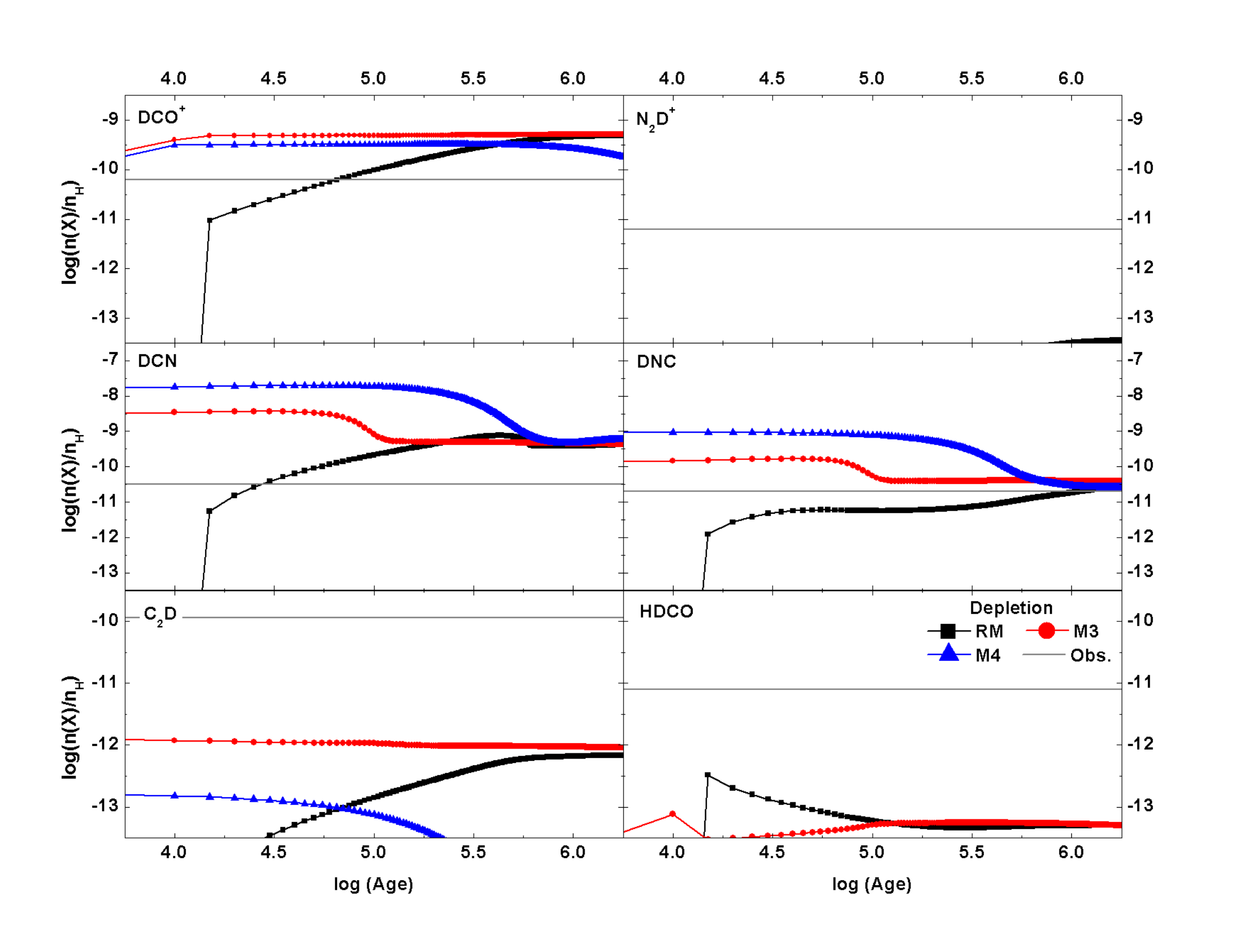}
\caption {The effect of varying the depletion of gaseous species onto grain surfaces on the time evolution of the calculated molecular abundances. Different curves represent different models: the RM (full depletion, black squares), model M3 (partial depletion, red circles), and model M4 (0\% depletion, blue triangles). Observations are represented by solid grey line as taken from \citet{wat15}.}
\label{fig:2}
\end{center}
\end{figure*}

Depletion is a measure of the amount of species removed from the gas to stick onto the grains. Therefore, changes in the depletion can be modelled by changing the sticking probability of the species onto grains \citep{raw92}. In our model, we followed this strategy in varying the depletion percentage and take the abundance of CO in the gas as the measure of the amount of depletion because it is the second most abundant gas-phase molecule  after H$_2$ (e.g. \citealt{viti04, awad10,awad14}).  
In the following we discuss the impact of variation in the depletion percentage of gaseous material onto grain surfaces on the fractional abundances of deuterated species. The amount of gaseous species involved in surface reactions was reduced 15\% of that amount in the RM in model M3 while the depletion was totally inhibited; i.e. become 0\% in model M4 to simulate pure gas-phase chemistry. 
Fig. \ref{fig:2} represents the time evolution of the deuterated species in CMM3 as calculated in cores with full depletion (RM, black squares), partial depletion (M3, red circles), and with no depletion (M4, blue triangles). The straight grey solid line is the observed abundance in CMM3 as taken from \citet{wat15}. 

When molecules are depleted onto grain surfaces, they get involved into surface reactions which enhances the abundance of their daughter molecules in the gas phase via desorption mechanisms. Therefore, by reducing the depletion onto grains we expect a general decrease in the abundances of the species because we are losing partially the contribution of surface chemistry. The results illustrated in Fig. \ref{fig:2} show that the sensitivity of species to depletion is different. The abundance of DCO$^+$, DCN, and DNC increases in models M3 (partial depletion) and model M4 (no depletion), in particular during early stages of the evolution (1.0 $\times$ 10$^4$ $<$ t(yr) $<$ 3.0 $\times$ 10$^{5}$) comparing to their abundances in the RM (full depletion). While the always underestimated species in our models (C$_2$D, HDCO, and N$_2$D$^+$) show also a respond to changes in the depletion onto grain surfaces. The abundance of N$_2$D$^+$ become invisible in both models M3 and M4, C$_2$D show an increase by 10 times that of the RM in both models specially at t $\le$ 3 $\times$ 10$^5$ yrs, and HDCO become barely visible in M3 and invisible in M4. These results may imply a grain origin either direct or indirect of these three species. It is remarkable that the abundances of these species either in models M3 or M4 always overestimate their observed values. 

The detailed analysis of the species DCO$^+$ and DCN showed that the main reason for their enhancement is the increase in their production rates in models M3 and M4 than that in the RM specially at early times. The reduction of the amount of species freezes onto grains increases their amount in the gas-phase allowing longer stay of the parent molecules in the region and hence more final products. In the case of DCN, the chemical analysis revealed a different chemistry at work in models M3 and M4 compared to the RM (Fig. \ref{fig:1.1}). The molecule is formed mainly by `C$_2$HD$^+$ + N' and the destruction in M3 happens by He$^+$ which is not as effective as H$^+$ in the RM which leads to more DCN in the core in model M3. In model M4, CR dissociation reactions are the main destruction routes of DCN, but also they are not as efficient as the destruction by H$^+$ in the RM so more DCN stayed in the gas. 

We found that the case of N$_2$D$^+$ is interesting because it does not show an enhancement in its abundance in contrary to other ions in the core. The analysis of this species showed that the ion N$_2$D$^+$ is formed in the gas-phase in the RM mainly by the reaction `orH$_2$D$^+$ + N$_2$' which become less efficient in model M3 and the formation took place via the deuterium exchange with N$_2$H$^+$, but with a formation rate 15 times less than that in the RM. In addition, the destruction of the ion by CO molecules become more efficient in model M3, 30 times higher than that in the RM because the abundance of CO in the former model is 15 - 20 times higher than the RM.

The increase in the amount of the deuterated species in partially depleted cores is in-line with observations of cold dense core \citep{bacm03}, however these models (M3 \& M4) cannot reproduce observations of CMM3 by \citet{wat15}. Thus we conclude that deuterated species are formed in fully depleted cores. 

\subsubsection{The CR Ionisation Rate}
\label{cr} 
\begin{figure*}
\begin{center} 
\includegraphics[trim= 1.5cm 0cm 1.0cm 0cm,clip=true,width=18cm]{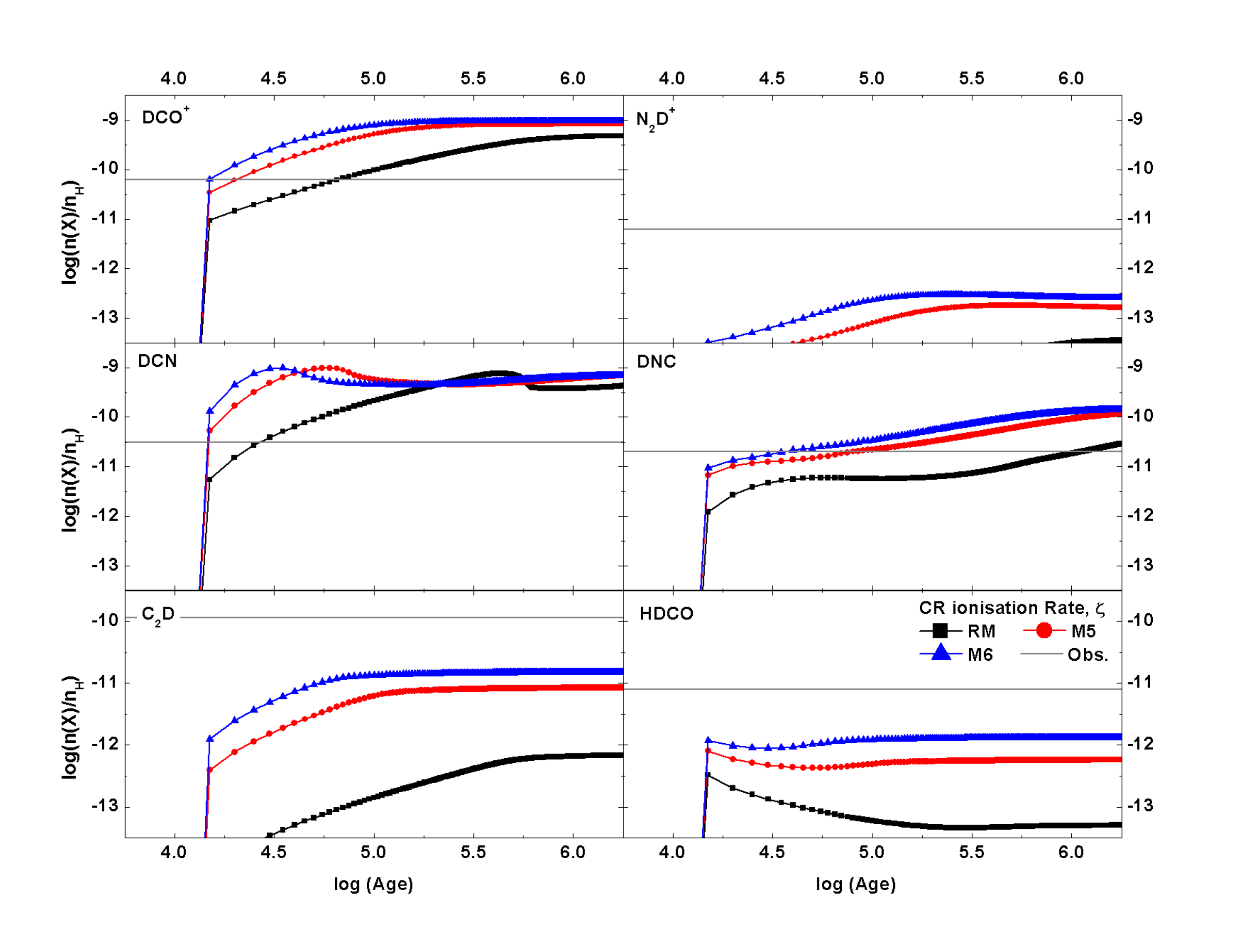}  
\caption {The chemical evolution of deuterated species as a function of time in cores with different CR ionisation rates, $\zeta$. The RM (1.3$\zeta_{\text {ISM}}$, black squares), model M5 (5$\zeta_{\text {ISM}}$, red circles) and model M6 (10$\zeta_{\text {ISM}}$, blue triangles)
Observed values, as taken from \citet{wat15}, are represented by solid grey line.}
\label{fig:3}
\end{center}
\end{figure*}

CR ionisation reactions are of special importance because they are the driver of the interstellar gas-phase chemistry since they are the main route of the formation of H$_3^+$ ions. When these ions react with HD molecules they form H$_2$D$^+$ which leads the interstellar deuterium gas chemistry \citep{mil89, mil05}. For this reason, we studied the influence of changing the CR ionisation rate, $\zeta$, on the abundances of deuterated species in the core. Several observations (e.g. \citealt{vant20, doty02, moro14, kaz15}) measured this rate in massive protostars and found it to be up to 500 times the standard interstellar value ($\zeta_{\text{ISM}}$ = 1.3 $\times$ 10$^{-17}$ s$^{-1}$; \citealt{lep92}) with a preferred range (5 - 10) $\zeta_{\text{ISM}}$. Most recently, \citet{awad17} calculated this rate (theoretically) for the massive protostellar core CMM3 and found that $\zeta_{\text{CMM3}}$ = 1.3 $\zeta_{\text{ISM}}$. In order to understand this influence in CMM3, we ran two models with higher ionisation rates than the RM; model M5 ($\zeta_{\text{CMM3}}$ = 5 $\zeta_{\text{ISM}}$) and model M6 ($\zeta_{\text{CMM3}}$ = 10 $\zeta_{\text{ISM}}$), see Table \ref{tab:grid}. The model results in comparison with the RM are illustrated in Fig. \ref{fig:3} in which results of models M5 and M6 are represented by red circles and blue triangles, respectively. From the figure, we notice that evolutionary trends show similarities to those obtained in the RM (black squares) but with higher abundances. The increase in the abundance is directly proportional to the increase in the ionisation rate. 

The chemical analysis of DCO$^+$ in models M5 and M6 showed similarities to that of the RM, as indicated by solid arrows in Fig. \ref{fig:1.1}. The formation of DCO$^+$ occurs by the reaction `orH$_2$D$^+$ + CO', at times t $\le$ 10$^4$ yrs, and the reaction `HCO$^+$ + D', for t $>$ 10$^4$ yrs. The enhancement of the CR ionisation rate enriches the medium with ions and hence increases the ultimate amount of DCO$^+$ (Fig. \ref{fig:1.1}). The simple molecule DCN is formed in models M5 and M6 by through the same route in the RM. Enhancing the CR ionisation rate, allow more dissociation reactions in the medium and enhanced the amount of radicals. This opened a new path for the production of the molecule which is `N + CHD' which increased the amount of DCN in the medium. 

It is noticeable that the results of models M5 and the RM are better match to observations than model M6 (blue line). From this we may conclude that the CR ionisation rate, $\zeta$, in the CMM3 is ranged between 1.3 - 5 $\zeta_{\text{ISM}}$. This result is in-line with both observations \citep{vant20} and previous models of CMM3 \citep{awad17}. 

\subsubsection{H$_2$ OPR}
\label{opr}
We ran few models similar in the physical conditions to the RM, but with different initial H$_2$ OPR; following the work by \citet{pag11}. We found that in the protostellar phase, the focus of this work, variations in the OPR do not affect the fractional abundances of deuterated species and hence their D/H ratio. On the other hand, deuterium chemistry found to be mostly affected by changes in the H$_2$ OPR during the dark cloud stage, which is the stage prior to the formation of the protostar. The chemical analysis of this dark cloud phase revealed that the abundances of deuterated species are enhanced by factors up to 10 when the OPR goes below 0.1, in particular for DCO$^+$ and N$_2$D$^+$. The chemistry of these two species is based on the abundances of the isotopologues of H$_3^+$ by their reactions with CO and N$_2$, respectively. The other studied molecules (DCN, DNC, C$_2$D, HDCO) do not show a significant change (factor $<$ 5) in their abundances with OPR because their main formation routes depends on CH$_2$D$^+$ and C$_2$HD$^+$ which are independent on OPR. In addition, these latter deuterium chemistry drivers are more important in regions with T $\ge$ 30K \citep{mil89}. In the protostellar phase and since the beginning of the chemistry, species tend to converge to values obtained in the RM and variations in the OPR do not have any influence on their abundances in particular at times $\ge$ 10$^4$ years when they become detectable. The reason for this might be the higher CR ionisation rate in CMM3 when compared to the dark cloud that in turn leads to destruction of the species during the very early times of the protostar. A comprehensive study on this topic and more observations may help for better understanding. 

\section{Model Limitations}
\label{lim}
Although our results for the performed models match observations for DCO$^+$, DCN and DNC at times $\le$ 10$^5$ yrs, under different physical conditions, all of our models failed in reproducing the observed amount of both C$_2$D and HDCO under any circumstances. 

The formation pathway of C$_2$D, as unveiled from the analysis of the RM, occurs in the gas-phase through the reaction of NH$_3$ with C$_2$HD$^+$ where the latter molecule is formed mainly by either reactions `HD + C$_2$H$_2$' or `C$^+$ + CH$_3$D'. The second pathway via deuterated methane (CH$_3$D) dominates the chemistry at times $>$ 10$^3$ yrs. Given the low temperature of the gas in CMM3 (15 K) and that the sublimation temperature of CH$_4$ ices is $\sim$ 30 K, the abundance of the reactants CH$_3$D and, therefore, C$_2$HD$^+$ will be very low and cannot account for the formation of larger amounts of C$_2$D. Another important point is that the deuterium enrichment via C$_2$HD$^+$ occurs in warm media \citep{mil89}, and hence the efficiency of deuteration for this molecule is going to be low. 

On the other hand, the formation of HDCO takes place both in the gas and on grain surfaces in which it is believed that the efficiency of mantle production is higher \citep{tiel83, wat08}. In the gas-phase, HDCO formation is based on the molecule C$_2$HD and CHD where both are daughters of CH$_2$D$^+$ and C$_2$HD$^+$. These two parents have low abundances in the gas, as calculated in our RM, (see bottom panel in Fig. \ref{fig:rm}) leading to the low obtained abundance of HDCO.

\begin{figure*}
\begin{center} 
\includegraphics[trim= 1.5cm 0cm 1.0cm 0cm,clip=true,width=18cm]{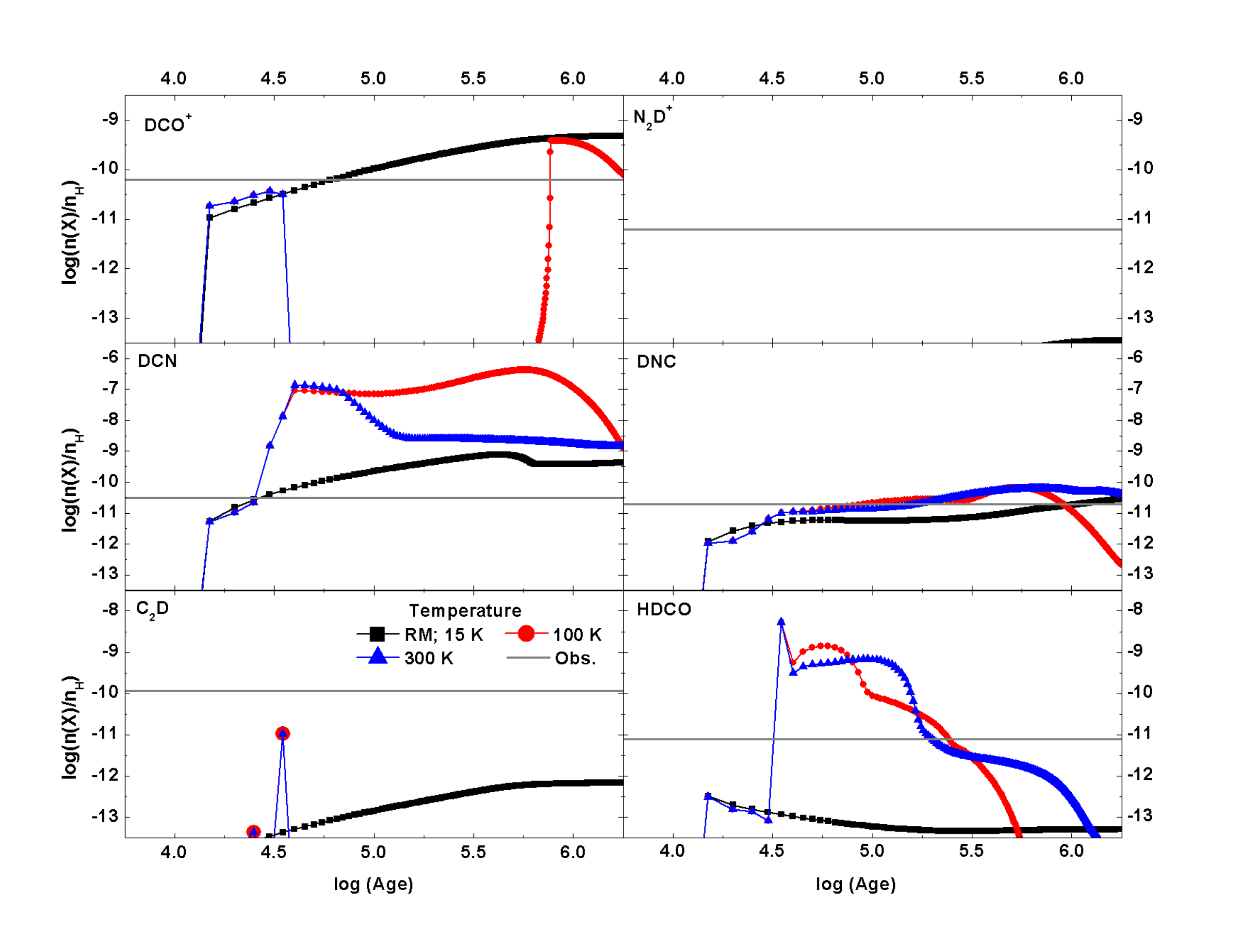}  
\caption {The chemical evolution of deuterated species as a function of time under higher temperatures T = 100 K (red circles), and 300 K (blue triangles) in comparison with the RM at 15 K (black squares). Observed values, as taken from \citet{wat15}, are represented by solid grey line.}
\label{fig:HC}
\end{center}
\end{figure*}

There is a possibility that HDCO and C$_2$D are formed in warmer regions (e.g. hot core; $\ge$ 100 K), rather than the cold (15 K) protostellar core observed by \citet{wat15}. In order to test this possibility, we ran a hot core model using the physical parameters of the protostellar core of CMM3, but at higher temperatures of 100 K and 300 K; see Fig. \ref{fig:HC}. The results of both models show that C$_2$D is enhanced at times correspond to the temperature range 70 - 90 K then it declines and become undetectable. The model calculations still underestimate the abundance of C$_2$D. The analysis showed that the decrease in C$_2$D abundance at times t $>$ 31,000 yrs is mainly due to (i) its destruction by gaseous O$_2$, and (ii) the destruction of its parent molecule, C$_2$HD$^+$, by para-H$_2$ which are very abundant in the core. On the contrary, the abundance of HDCO increases and exceeds the observed value, at times when T $\ge$ 100 K (for both models) due to its sublimation from grains. Surprisingly, DNC showed a better agreement with observations at the hot core temperatures. These results may imply that HDCO and DNC may be formed in warmer areas than the other species and could be used as tracers of these hot core regions while models should consider another mechanism to account for warm carbon-chain molecules (e.g. C$_2$D) which is formed in warm gas but not necessarily in the hot core.

The amount of mantle C$_2$D and HDCO molecules are found to be comparable to their observed gaseous values by \citet{wat15}. This may lead to another limitation of the model, if these species are not formed in hot cores, which is the limited number of desorption mechanisms we use in our model at low temperatures. At the moment, we are working on including more non-thermal desorption mechanisms such as chemical explosions and sputtering which are important specially in dense regions. In addition, \citet{sar11} suggested that the core CMM3 might be exposed to or affected by slow shocks from its surrounding region. We are also considering studying the influence of including shocks on the chemistry of the region (Awad et al. in prep.). 

\section{Conclusions}
\label{conc} 
We studied the sensitivity of deuterium chemistry in the massive young protostellar core NGC 2264 CMM3 to its surrounding physical conditions by running gas-grain chemical models. We summarise our conclusions as follows:
\begin{enumerate}
\item [{\bf 1-}] In CMM3, the deuterium chemistry is sensitive to changes in the core density, depletion percentage, and the CR ionisation rate. Therefore, we can not use any of these studied molecules as a tracer to the physics of the region.
\item [{\bf 2-}] Deuterium chemistry is insensitive to changes in the H$_2$ OPR under the physical conditions of prestellar objects, but it is very sensitive to these variations in the pre-stage of dark clouds.
\item [{\bf 3-}] The model reproduces observations for cores with density range (1 - 5) $\times$ 10$^6$ cm$^{-3}$ or CR ionisation rate between (1.69 and 6.5) $\times$ 10$^{-17}$ s$^{-1}$. 
\item [{\bf 4-}] Time of best fit with observations (t $<$ 10$^5$ yrs) is in-line with the estimated age of young protostellar cores \citep{and93, and20}. We then confirms that CMM3 is in its early evolutionary stage and its estimated age is (1 - 5) $\times$ 10$^4$ yrs.
\item [{\bf 5-}] HDCO and DNC might be formed in hot cores rather than the cold protostellar region and hence they may trace hotter gas ($>$ 100K) while the formation of C$_2$D is linked to warm deuterium chemistry, but they are not necessarily trace the hot core region.
\end{enumerate}

\section*{Acknowledgements}
The authors would like to thank the referee for the valuable comments that helped in improving the original manuscript.


\begin{thebibliography}{73}
\ifx \bisbn   \undefined \def \bisbn  #1{ISBN #1}\fi
\ifx \binits  \undefined \def \binits#1{#1} \fi
\ifx \bauthor  \undefined \def \bauthor#1{#1} \fi
\ifx \batitle  \undefined \def \batitle#1{#1} \fi
\ifx \bjtitle  \undefined \def \bjtitle#1{#1}\fi
\ifx \bvolume  \undefined \def \bvolume#1{\textbf{#1}}\fi
\ifx \byear  \undefined \def \byear#1{#1} \fi
\ifx \bissue  \undefined \def \bissue#1{#1} \fi
\ifx \bfpage  \undefined \def \bfpage#1{#1} \fi
\ifx \blpage  \undefined \def \blpage #1{#1} \fi
\ifx \burl  \undefined \def \burl#1{\textsf{#1}} \fi
\ifx \doiurl  \undefined \def \doiurl#1{\textsf{#1}} \fi
\ifx \betal  \undefined \def \betal{\textit{et al.}} \fi
\ifx \binstitute  \undefined \def \binstitute#1{#1} \fi
\ifx \binstitutionaled  \undefined \def \binstitutionaled#1{#1} \fi
\ifx \bctitle  \undefined \def \bctitle#1{#1} \fi
\ifx \beditor  \undefined \def \beditor#1{#1} \fi
\ifx \bpublisher  \undefined \def \bpublisher#1{#1} \fi
\ifx \bbtitle  \undefined \def \bbtitle#1{#1} \fi
\ifx \bedition  \undefined \def \bedition#1{#1} \fi
\ifx \bseriesno  \undefined \def \bseriesno#1{#1} \fi
\ifx \blocation  \undefined \def \blocation#1{#1} \fi
\ifx \bsertitle  \undefined \def \bsertitle#1{#1} \fi
\ifx \bsnm \undefined \def \bsnm#1{#1} \fi
\ifx \bsuffix \undefined \def \bsuffix#1{#1} \fi
\ifx \bparticle \undefined \def \bparticle#1{#1} \fi
\ifx \barticle \undefined \def \barticle#1{#1} \fi
\ifx \bconfdate \undefined \def \bconfdate #1{#1} \fi
\ifx \botherref \undefined \def \botherref #1{#1} \fi
\ifx \url \undefined \def \url#1{\textsf{#1}} \fi
\ifx \bchapter \undefined \def \bchapter#1{#1} \fi
\ifx \bbook \undefined \def \bbook#1{#1} \fi
\ifx \bcomment \undefined \def \bcomment#1{#1} \fi
\ifx \oauthor \undefined \def \oauthor#1{#1} \fi
\ifx \citeauthoryear \undefined \def \citeauthoryear#1{#1} \fi
\ifx \endbibitem  \undefined \def \endbibitem {}\fi
\ifx \bconflocation  \undefined \def \bconflocation#1{#1} \fi
\ifx \arxivurl  \undefined \def \arxivurl#1{\textsf{#1}} \fi

\bibitem[\protect\citeauthoryear{{Aikawa} et~al.}{2012}]{aik12}
\begin{barticle}
\bauthor{\bsnm{{Aikawa}}, \binits{Y.}},
\bauthor{\bsnm{{Wakelam}}, \binits{V.}},
\bauthor{\bsnm{{Hersant}}, \binits{F.}},
\bauthor{\bsnm{{Garrod}}, \binits{R.T.}},
\bauthor{\bsnm{{Herbst}}, \binits{E.}}:
\bjtitle{\apj}
\bvolume{760},
\bfpage{40}
(\byear{2012}).
\arxivurl{1210.2476}.
doi:\doiurl{10.1088/0004-637X/760/1/40}
\end{barticle}
\endbibitem

\bibitem[\protect\citeauthoryear{{Albertsson} et~al.}{2013}]{albrt013}
\begin{barticle}
\bauthor{\bsnm{{Albertsson}}, \binits{T.}},
\bauthor{\bsnm{{Semenov}}, \binits{D.A.}},
\bauthor{\bsnm{{Vasyunin}}, \binits{A.I.}},
\bauthor{\bsnm{{Henning}}, \binits{T.}},
\bauthor{\bsnm{{Herbst}}, \binits{E.}}:
\bjtitle{\apjs}
\bvolume{207},
\bfpage{27}
(\byear{2013}).
doi:\doiurl{10.1088/0067-0049/207/2/27}
\end{barticle}
\endbibitem

\bibitem[\protect\citeauthoryear{{Andr\'{e}} et~al.}{1993}]{and93}
\begin{barticle}
\bauthor{\bsnm{{Andr\'{e}}}, \binits{P.}},
\bauthor{\bsnm{{Ward-Thompson}}, \binits{D.}},
\bauthor{\bsnm{{Barsony}}, \binits{M.}}:
\bjtitle{\apj}
\bvolume{406},
\bfpage{122}
(\byear{1993}).
doi:\doiurl{10.1086/172425}
\end{barticle}
\endbibitem

\bibitem[\protect\citeauthoryear{{Andr\'{e}} et~al.}{2000}]{and20}
\begin{botherref}
\oauthor{\bsnm{{Andr\'{e}}}, \binits{P.}},
\oauthor{\bsnm{{Ward-Thompson}}, \binits{D.}},
\oauthor{\bsnm{{Barsony}}, \binits{M.}}:
Protostars and Planets IV,
59
(2000).
\arxivurl{arXiv:astro-ph/9903284}
\end{botherref}
\endbibitem

\bibitem[\protect\citeauthoryear{{Asplund} et~al.}{2009}]{asp09}
\begin{barticle}
\bauthor{\bsnm{{Asplund}}, \binits{M.}},
\bauthor{\bsnm{{Grevesse}}, \binits{N.}},
\bauthor{\bsnm{{Sauval}}, \binits{A.J.}},
\bauthor{\bsnm{{Scott}}, \binits{P.}}:
\bjtitle{\araa}
\bvolume{47},
\bfpage{481}
(\byear{2009}).
\arxivurl{0909.0948}.
doi:\doiurl{10.1146/annurev.astro.46.060407.145222}
\end{barticle}
\endbibitem

\bibitem[\protect\citeauthoryear{{Awad} and {Shalabeia}}{2017}]{awad17}
\begin{barticle}
\bauthor{\bsnm{{Awad}}, \binits{Z.}},
\bauthor{\bsnm{{Shalabeia}}, \binits{O.M.}}:
\bjtitle{\apss}
\bvolume{362},
\bfpage{83}
(\byear{2017}).
\arxivurl{1703.04632}.
doi:\doiurl{10.1007/s10509-017-3061-8}
\end{barticle}
\endbibitem

\bibitem[\protect\citeauthoryear{{Awad} et~al.}{2010}]{awad10}
\begin{botherref}
\oauthor{\bsnm{{Awad}}, \binits{Z.}},
\oauthor{\bsnm{{Viti}}, \binits{S.}},
\oauthor{\bsnm{{Collings}}, \binits{M.P.}},
\oauthor{\bsnm{{Williams}}, \binits{D.A.}}:
\mnras,
1006
(2010).
\arxivurl{1005.5265}.
doi:\doiurl{10.1111/j.1365-2966.2010.17077.x}
\end{botherref}
\endbibitem

\bibitem[\protect\citeauthoryear{{Awad} et~al.}{2014}]{awad14}
\begin{barticle}
\bauthor{\bsnm{{Awad}}, \binits{Z.}},
\bauthor{\bsnm{{Viti}}, \binits{S.}},
\bauthor{\bsnm{{Bayet}}, \binits{E.}},
\bauthor{\bsnm{{Caselli}}, \binits{P.}}:
\bjtitle{\mnras}
\bvolume{443},
\bfpage{275}
(\byear{2014}).
\arxivurl{1406.2272}.
doi:\doiurl{10.1093/mnras/stu1141}
\end{barticle}
\endbibitem

\bibitem[\protect\citeauthoryear{{Bacmann} et~al.}{2003}]{bacm03}
\begin{barticle}
\bauthor{\bsnm{{Bacmann}}, \binits{A.}},
\bauthor{\bsnm{{Lefloch}}, \binits{B.}},
\bauthor{\bsnm{{Ceccarelli}}, \binits{C.}},
\bauthor{\bsnm{{Steinacker}}, \binits{J.}},
\bauthor{\bsnm{{Castets}}, \binits{A.}},
\bauthor{\bsnm{{Loinard}}, \binits{L.}}:
\bjtitle{\apjl}
\bvolume{585},
\bfpage{55}
(\byear{2003}).
\arxivurl{arXiv:astro-ph/0301651}.
doi:\doiurl{10.1086/374263}
\end{barticle}
\endbibitem

\bibitem[\protect\citeauthoryear{{Bergman} et~al.}{2011}]{berg11}
\begin{barticle}
\bauthor{\bsnm{{Bergman}}, \binits{P.}},
\bauthor{\bsnm{{Parise}}, \binits{B.}},
\bauthor{\bsnm{{Liseau}}, \binits{R.}},
\bauthor{\bsnm{{Larsson}}, \binits{B.}}:
\bjtitle{\aap}
\bvolume{527},
\bfpage{39}
(\byear{2011}).
\arxivurl{1011.3339}.
doi:\doiurl{10.1051/0004-6361/201015012}
\end{barticle}
\endbibitem

\bibitem[\protect\citeauthoryear{{Brown} and {Millar}}{1989a}]{bro89b}
\begin{barticle}
\bauthor{\bsnm{{Brown}}, \binits{P.D.}},
\bauthor{\bsnm{{Millar}}, \binits{T.J.}}:
\bjtitle{\mnras}
\bvolume{240},
\bfpage{25}
(\byear{1989}a)
\end{barticle}
\endbibitem

\bibitem[\protect\citeauthoryear{{Brown} and {Millar}}{1989b}]{bro89a}
\begin{barticle}
\bauthor{\bsnm{{Brown}}, \binits{P.D.}},
\bauthor{\bsnm{{Millar}}, \binits{T.J.}}:
\bjtitle{\mnras}
\bvolume{237},
\bfpage{661}
(\byear{1989}b)
\end{barticle}
\endbibitem

\bibitem[\protect\citeauthoryear{{Caselli} and {Ceccarelli}}{2012}]{cas012}
\begin{barticle}
\bauthor{\bsnm{{Caselli}}, \binits{P.}},
\bauthor{\bsnm{{Ceccarelli}}, \binits{C.}}:
\bjtitle{\aapr}
\bvolume{20},
\bfpage{56}
(\byear{2012}).
\arxivurl{1210.6368}.
doi:\doiurl{10.1007/s00159-012-0056-x}
\end{barticle}
\endbibitem

\bibitem[\protect\citeauthoryear{{Ceccarelli} et~al.}{2001}]{cec01}
\begin{barticle}
\bauthor{\bsnm{{Ceccarelli}}, \binits{C.}},
\bauthor{\bsnm{{Loinard}}, \binits{L.}},
\bauthor{\bsnm{{Castets}}, \binits{A.}},
\bauthor{\bsnm{{Tielens}}, \binits{A.G.G.M.}},
\bauthor{\bsnm{{Caux}}, \binits{E.}},
\bauthor{\bsnm{{Lefloch}}, \binits{B.}},
\bauthor{\bsnm{{Vastel}}, \binits{C.}}:
\bjtitle{\aap}
\bvolume{372},
\bfpage{998}
(\byear{2001}).
doi:\doiurl{10.1051/0004-6361:20010559}
\end{barticle}
\endbibitem

\bibitem[\protect\citeauthoryear{{Ceccarelli} et~al.}{2007}]{cec07}
\begin{bchapter}
\bauthor{\bsnm{{Ceccarelli}}, \binits{C.}},
\bauthor{\bsnm{{Caselli}}, \binits{P.}},
\bauthor{\bsnm{{Herbst}}, \binits{E.}},
\bauthor{\bsnm{{Tielens}}, \binits{A.G.G.M.}},
\bauthor{\bsnm{{Caux}}, \binits{E.}}:
In: \beditor{\bsnm{{Reipurth}}, \binits{B.}},
\beditor{\bsnm{{Jewitt}}, \binits{D.}},
\beditor{\bsnm{{Keil}}, \binits{K.}} (eds.)
\bbtitle{Protostars and Planets V},
p. \bfpage{47}
(\byear{2007})
\end{bchapter}
\endbibitem

\bibitem[\protect\citeauthoryear{{Coutens} et~al.}{2013}]{cout13}
\begin{barticle}
\bauthor{\bsnm{{Coutens}}, \binits{A.}},
\bauthor{\bsnm{{Vastel}}, \binits{C.}},
\bauthor{\bsnm{{Cazaux}}, \binits{S.}},
\bauthor{\bsnm{{Bottinelli}}, \binits{S.}},
\bauthor{\bsnm{{Caux}}, \binits{E.}},
\bauthor{\bsnm{{Ceccarelli}}, \binits{C.}},
\bauthor{\bsnm{{Demyk}}, \binits{K.}},
\bauthor{\bsnm{{Taquet}}, \binits{V.}},
\bauthor{\bsnm{{Wakelam}}, \binits{V.}}:
\bjtitle{\aap}
\bvolume{553},
\bfpage{75}
(\byear{2013}).
\arxivurl{1304.2890}.
doi:\doiurl{10.1051/0004-6361/201220967}
\end{barticle}
\endbibitem

\bibitem[\protect\citeauthoryear{{Coutens} et~al.}{2014}]{cout14}
\begin{barticle}
\bauthor{\bsnm{{Coutens}}, \binits{A.}},
\bauthor{\bsnm{{Vastel}}, \binits{C.}},
\bauthor{\bsnm{{Hincelin}}, \binits{U.}},
\bauthor{\bsnm{{Herbst}}, \binits{E.}},
\bauthor{\bsnm{{Lis}}, \binits{D.C.}},
\bauthor{\bsnm{{Chavarr{\'{\i}}a}}, \binits{L.}},
\bauthor{\bsnm{{G{\'e}rin}}, \binits{M.}},
\bauthor{\bsnm{{van der Tak}}, \binits{F.F.S.}},
\bauthor{\bsnm{{Persson}}, \binits{C.M.}},
\bauthor{\bsnm{{Goldsmith}}, \binits{P.F.}},
\bauthor{\bsnm{{Caux}}, \binits{E.}}:
\bjtitle{\mnras}
\bvolume{445},
\bfpage{1299}
(\byear{2014}).
\arxivurl{1409.1092}.
doi:\doiurl{10.1093/mnras/stu1816}
\end{barticle}
\endbibitem

\bibitem[\protect\citeauthoryear{{Crapsi} et~al.}{2005}]{crap05}
\begin{barticle}
\bauthor{\bsnm{{Crapsi}}, \binits{A.}},
\bauthor{\bsnm{{Caselli}}, \binits{P.}},
\bauthor{\bsnm{{Walmsley}}, \binits{C.M.}},
\bauthor{\bsnm{{Myers}}, \binits{P.C.}},
\bauthor{\bsnm{{Tafalla}}, \binits{M.}},
\bauthor{\bsnm{{Lee}}, \binits{C.W.}},
\bauthor{\bsnm{{Bourke}}, \binits{T.L.}}:
\bjtitle{\apj}
\bvolume{619},
\bfpage{379}
(\byear{2005}).
\arxivurl{arXiv:astro-ph/0409529}.
doi:\doiurl{10.1086/426472}
\end{barticle}
\endbibitem

\bibitem[\protect\citeauthoryear{{Cuppen} et~al.}{2017}]{cup17}
\begin{barticle}
\bauthor{\bsnm{{Cuppen}}, \binits{H.M.}},
\bauthor{\bsnm{{Walsh}}, \binits{C.}},
\bauthor{\bsnm{{Lamberts}}, \binits{T.}},
\bauthor{\bsnm{{Semenov}}, \binits{D.}},
\bauthor{\bsnm{{Garrod}}, \binits{R.T.}},
\bauthor{\bsnm{{Penteado}}, \binits{E.M.}},
\bauthor{\bsnm{{Ioppolo}}, \binits{S.}}:
\bjtitle{\ssr}
\bvolume{212},
\bfpage{1}
(\byear{2017}).
doi:\doiurl{10.1007/s11214-016-0319-3}
\end{barticle}
\endbibitem

\bibitem[\protect\citeauthoryear{{Doty} et~al.}{2002}]{doty02}
\begin{barticle}
\bauthor{\bsnm{{Doty}}, \binits{S.D.}},
\bauthor{\bsnm{{van Dishoeck}}, \binits{E.F.}},
\bauthor{\bsnm{{van der Tak}}, \binits{F.F.S.}},
\bauthor{\bsnm{{Boonman}}, \binits{A.M.S.}}:
\bjtitle{\aap}
\bvolume{389},
\bfpage{446}
(\byear{2002}).
\arxivurl{astro-ph/0205292}.
doi:\doiurl{10.1051/0004-6361:20020597}
\end{barticle}
\endbibitem

\bibitem[\protect\citeauthoryear{{Flower} et~al.}{2006}]{flo06}
\begin{barticle}
\bauthor{\bsnm{{Flower}}, \binits{D.R.}},
\bauthor{\bsnm{{Pineau Des For{\^e}ts}}, \binits{G.}},
\bauthor{\bsnm{{Walmsley}}, \binits{C.M.}}:
\bjtitle{\aap}
\bvolume{449},
\bfpage{621}
(\byear{2006}).
\arxivurl{arXiv:astro-ph/0601429}.
doi:\doiurl{10.1051/0004-6361:20054246}
\end{barticle}
\endbibitem

\bibitem[\protect\citeauthoryear{{Furuya} et~al.}{2016}]{fur16}
\begin{barticle}
\bauthor{\bsnm{{Furuya}}, \binits{K.}},
\bauthor{\bsnm{{van Dishoeck}}, \binits{E.F.}},
\bauthor{\bsnm{{Aikawa}}, \binits{Y.}}:
\bjtitle{\aap}
\bvolume{586},
\bfpage{127}
(\byear{2016}).
\arxivurl{1512.04291}.
doi:\doiurl{10.1051/0004-6361/201527579}
\end{barticle}
\endbibitem

\bibitem[\protect\citeauthoryear{{Furuya} et~al.}{2013}]{fur13}
\begin{barticle}
\bauthor{\bsnm{{Furuya}}, \binits{K.}},
\bauthor{\bsnm{{Aikawa}}, \binits{Y.}},
\bauthor{\bsnm{{Nomura}}, \binits{H.}},
\bauthor{\bsnm{{Hersant}}, \binits{F.}},
\bauthor{\bsnm{{Wakelam}}, \binits{V.}}:
\bjtitle{\apj}
\bvolume{779},
\bfpage{11}
(\byear{2013}).
\arxivurl{1310.3342}.
doi:\doiurl{10.1088/0004-637X/779/1/11}
\end{barticle}
\endbibitem

\bibitem[\protect\citeauthoryear{{Gerlich} et~al.}{2002}]{gerl02}
\begin{barticle}
\bauthor{\bsnm{{Gerlich}}, \binits{D.}},
\bauthor{\bsnm{{Herbst}}, \binits{E.}},
\bauthor{\bsnm{{Roueff}}, \binits{E.}}:
\bjtitle{\planss}
\bvolume{50},
\bfpage{1275}
(\byear{2002}).
doi:\doiurl{10.1016/S0032-0633(02)00094-6}
\end{barticle}
\endbibitem

\bibitem[\protect\citeauthoryear{{Herbst} and {van Dishoeck}}{2009}]{her09}
\begin{barticle}
\bauthor{\bsnm{{Herbst}}, \binits{E.}},
\bauthor{\bsnm{{van Dishoeck}}, \binits{E.F.}}:
\bjtitle{\araa}
\bvolume{47},
\bfpage{427}
(\byear{2009}).
doi:\doiurl{10.1146/annurev-astro-082708-101654}
\end{barticle}
\endbibitem

\bibitem[\protect\citeauthoryear{{Hincelin} et~al.}{2014}]{hin14}
\begin{bchapter}
\bauthor{\bsnm{{Hincelin}}, \binits{U.}},
\bauthor{\bsnm{{Herbst}}, \binits{E.}},
\bauthor{\bsnm{{Chang}}, \binits{Q.}},
\bauthor{\bsnm{{Vasyunina}}, \binits{T.}},
\bauthor{\bsnm{{Aikawa}}, \binits{Y.}},
\bauthor{\bsnm{{Furuya}}, \binits{K.}}:
In: \bbtitle{69th International Symposium on Molecular Spectroscopy},
p. \bfpage{9}
(\byear{2014})
\end{bchapter}
\endbibitem

\bibitem[\protect\citeauthoryear{{Huang} and {{\"O}berg}}{2015}]{hua15}
\begin{barticle}
\bauthor{\bsnm{{Huang}}, \binits{J.}},
\bauthor{\bsnm{{{\"O}berg}}, \binits{K.I.}}:
\bjtitle{\apjl}
\bvolume{809},
\bfpage{26}
(\byear{2015}).
\arxivurl{1508.03637}.
doi:\doiurl{10.1088/2041-8205/809/2/L26}
\end{barticle}
\endbibitem

\bibitem[\protect\citeauthoryear{{Ka{\'z}mierczak-Barthel}
  et~al.}{2015}]{kaz15}
\begin{barticle}
\bauthor{\bsnm{{Ka{\'z}mierczak-Barthel}}, \binits{M.}},
\bauthor{\bsnm{{Semenov}}, \binits{D.A.}},
\bauthor{\bsnm{{van der Tak}}, \binits{F.F.S.}},
\bauthor{\bsnm{{Chavarr{\'{\i}}a}}, \binits{L.}},
\bauthor{\bsnm{{van der Wiel}}, \binits{M.H.D.}}:
\bjtitle{\aap}
\bvolume{574},
\bfpage{71}
(\byear{2015}).
\arxivurl{1412.5763}.
doi:\doiurl{10.1051/0004-6361/201424657}
\end{barticle}
\endbibitem

\bibitem[\protect\citeauthoryear{{Lepp}}{1992}]{lep92}
\begin{bchapter}
\bauthor{\bsnm{{Lepp}}, \binits{S.}}:
In: \beditor{\bsnm{{Singh}}, \binits{P.D.}} (ed.)
\bbtitle{Astrochemistry of Cosmic Phenomena}.
\bsertitle{IAU Symposium},
vol. \bseriesno{150},
p. \bfpage{471}
(\byear{1992})
\end{bchapter}
\endbibitem

\bibitem[\protect\citeauthoryear{{Loinard} et~al.}{2001}]{loi01}
\begin{barticle}
\bauthor{\bsnm{{Loinard}}, \binits{L.}},
\bauthor{\bsnm{{Castets}}, \binits{A.}},
\bauthor{\bsnm{{Ceccarelli}}, \binits{C.}},
\bauthor{\bsnm{{Caux}}, \binits{E.}},
\bauthor{\bsnm{{Tielens}}, \binits{A.G.G.M.}}:
\bjtitle{\apjl}
\bvolume{552},
\bfpage{163}
(\byear{2001}).
doi:\doiurl{10.1086/320331}
\end{barticle}
\endbibitem

\bibitem[\protect\citeauthoryear{{Loren} and {Wootten}}{1985}]{lor85}
\begin{barticle}
\bauthor{\bsnm{{Loren}}, \binits{R.B.}},
\bauthor{\bsnm{{Wootten}}, \binits{A.}}:
\bjtitle{\apj}
\bvolume{299},
\bfpage{947}
(\byear{1985}).
doi:\doiurl{10.1086/163761}
\end{barticle}
\endbibitem

\bibitem[\protect\citeauthoryear{{Majumdar} et~al.}{2017}]{maj17}
\begin{barticle}
\bauthor{\bsnm{{Majumdar}}, \binits{L.}},
\bauthor{\bsnm{{Gratier}}, \binits{P.}},
\bauthor{\bsnm{{Ruaud}}, \binits{M.}},
\bauthor{\bsnm{{Wakelam}}, \binits{V.}},
\bauthor{\bsnm{{Vastel}}, \binits{C.}},
\bauthor{\bsnm{{Sipil{\"a}}}, \binits{O.}},
\bauthor{\bsnm{{Hersant}}, \binits{F.}},
\bauthor{\bsnm{{Dutrey}}, \binits{A.}},
\bauthor{\bsnm{{Guilloteau}}, \binits{S.}}:
\bjtitle{\mnras}
\bvolume{466},
\bfpage{4470}
(\byear{2017}).
\arxivurl{1612.07845}.
doi:\doiurl{10.1093/mnras/stw3360}
\end{barticle}
\endbibitem

\bibitem[\protect\citeauthoryear{{Maury} et~al.}{2009}]{mau09}
\begin{barticle}
\bauthor{\bsnm{{Maury}}, \binits{A.J.}},
\bauthor{\bsnm{{Andr{\'e}}}, \binits{P.}},
\bauthor{\bsnm{{Li}}, \binits{Z.-Y.}}:
\bjtitle{\aap}
\bvolume{499},
\bfpage{175}
(\byear{2009}).
\arxivurl{0902.1379}.
doi:\doiurl{10.1051/0004-6361/200811442}
\end{barticle}
\endbibitem

\bibitem[\protect\citeauthoryear{{McElroy} et~al.}{2013}]{mce013}
\begin{barticle}
\bauthor{\bsnm{{McElroy}}, \binits{D.}},
\bauthor{\bsnm{{Walsh}}, \binits{C.}},
\bauthor{\bsnm{{Markwick}}, \binits{A.J.}},
\bauthor{\bsnm{{Cordiner}}, \binits{M.A.}},
\bauthor{\bsnm{{Smith}}, \binits{K.}},
\bauthor{\bsnm{{Millar}}, \binits{T.J.}}:
\bjtitle{\aap}
\bvolume{550},
\bfpage{36}
(\byear{2013}).
\arxivurl{1212.6362}.
doi:\doiurl{10.1051/0004-6361/201220465}
\end{barticle}
\endbibitem

\bibitem[\protect\citeauthoryear{{Miettinen} et~al.}{2012}]{mie12}
\begin{barticle}
\bauthor{\bsnm{{Miettinen}}, \binits{O.}},
\bauthor{\bsnm{{Harju}}, \binits{J.}},
\bauthor{\bsnm{{Haikala}}, \binits{L.K.}},
\bauthor{\bsnm{{Juvela}}, \binits{M.}}:
\bjtitle{\aap}
\bvolume{538},
\bfpage{137}
(\byear{2012}).
\arxivurl{1112.5053}.
doi:\doiurl{10.1051/0004-6361/201117849}
\end{barticle}
\endbibitem

\bibitem[\protect\citeauthoryear{{Millar}}{2005}]{mil05}
\begin{barticle}
\bauthor{\bsnm{{Millar}}, \binits{T.J.}}:
\bjtitle{Astronomy and Geophysics}
\bvolume{46}(\bissue{2}),
\bfpage{020000}
(\byear{2005}).
doi:\doiurl{10.1111/j.1468-4004.2005.46229.x}
\end{barticle}
\endbibitem

\bibitem[\protect\citeauthoryear{{Millar} et~al.}{1989}]{mil89}
\begin{barticle}
\bauthor{\bsnm{{Millar}}, \binits{T.J.}},
\bauthor{\bsnm{{Bennett}}, \binits{A.}},
\bauthor{\bsnm{{Herbst}}, \binits{E.}}:
\bjtitle{\apj}
\bvolume{340},
\bfpage{906}
(\byear{1989}).
doi:\doiurl{10.1086/167444}
\end{barticle}
\endbibitem

\bibitem[\protect\citeauthoryear{{Millar} et~al.}{1997}]{mil97}
\begin{barticle}
\bauthor{\bsnm{{Millar}}, \binits{T.J.}},
\bauthor{\bsnm{{Farquhar}}, \binits{P.R.A.}},
\bauthor{\bsnm{{Willacy}}, \binits{K.}}:
\bjtitle{\aaps}
\bvolume{121},
\bfpage{139}
(\byear{1997}).
doi:\doiurl{10.1051/aas:1997118}
\end{barticle}
\endbibitem

\bibitem[\protect\citeauthoryear{{Morales Ortiz} et~al.}{2014}]{moro14}
\begin{barticle}
\bauthor{\bsnm{{Morales Ortiz}}, \binits{J.L.}},
\bauthor{\bsnm{{Ceccarelli}}, \binits{C.}},
\bauthor{\bsnm{{Lis}}, \binits{D.C.}},
\bauthor{\bsnm{{Olmi}}, \binits{L.}},
\bauthor{\bsnm{{Plume}}, \binits{R.}},
\bauthor{\bsnm{{Schilke}}, \binits{P.}}:
\bjtitle{\aap}
\bvolume{563},
\bfpage{127}
(\byear{2014}).
\arxivurl{1306.3012}.
doi:\doiurl{10.1051/0004-6361/201322071}
\end{barticle}
\endbibitem

\bibitem[\protect\citeauthoryear{{{\"O}berg} et~al.}{2015}]{ob15}
\begin{barticle}
\bauthor{\bsnm{{{\"O}berg}}, \binits{K.I.}},
\bauthor{\bsnm{{Furuya}}, \binits{K.}},
\bauthor{\bsnm{{Loomis}}, \binits{R.}},
\bauthor{\bsnm{{Aikawa}}, \binits{Y.}},
\bauthor{\bsnm{{Andrews}}, \binits{S.M.}},
\bauthor{\bsnm{{Qi}}, \binits{C.}},
\bauthor{\bsnm{{van Dishoeck}}, \binits{E.F.}},
\bauthor{\bsnm{{Wilner}}, \binits{D.J.}}:
\bjtitle{\apj}
\bvolume{810},
\bfpage{112}
(\byear{2015}).
\arxivurl{1508.07296}.
doi:\doiurl{10.1088/0004-637X/810/2/112}
\end{barticle}
\endbibitem

\bibitem[\protect\citeauthoryear{{Oliveira} et~al.}{2003}]{oli03}
\begin{barticle}
\bauthor{\bsnm{{Oliveira}}, \binits{C.M.}},
\bauthor{\bsnm{{H{\'e}brard}}, \binits{G.}},
\bauthor{\bsnm{{Howk}}, \binits{J.C.}},
\bauthor{\bsnm{{Kruk}}, \binits{J.W.}},
\bauthor{\bsnm{{Chayer}}, \binits{P.}},
\bauthor{\bsnm{{Moos}}, \binits{H.W.}}:
\bjtitle{\apj}
\bvolume{587},
\bfpage{235}
(\byear{2003}).
\arxivurl{arXiv:astro-ph/0212506}.
doi:\doiurl{10.1086/368019}
\end{barticle}
\endbibitem

\bibitem[\protect\citeauthoryear{{Pagani} et~al.}{2011}]{pag11}
\begin{barticle}
\bauthor{\bsnm{{Pagani}}, \binits{L.}},
\bauthor{\bsnm{{Roueff}}, \binits{E.}},
\bauthor{\bsnm{{Lesaffre}}, \binits{P.}}:
\bjtitle{\apjl}
\bvolume{739},
\bfpage{35}
(\byear{2011}).
\arxivurl{1109.6495}.
doi:\doiurl{10.1088/2041-8205/739/2/L35}
\end{barticle}
\endbibitem

\bibitem[\protect\citeauthoryear{{Peretto} et~al.}{2006}]{per06}
\begin{barticle}
\bauthor{\bsnm{{Peretto}}, \binits{N.}},
\bauthor{\bsnm{{Andr{\'e}}}, \binits{P.}},
\bauthor{\bsnm{{Belloche}}, \binits{A.}}:
\bjtitle{\aap}
\bvolume{445},
\bfpage{979}
(\byear{2006}).
\arxivurl{astro-ph/0508619}.
doi:\doiurl{10.1051/0004-6361:20053324}
\end{barticle}
\endbibitem

\bibitem[\protect\citeauthoryear{{Peretto} et~al.}{2007}]{per07}
\begin{barticle}
\bauthor{\bsnm{{Peretto}}, \binits{N.}},
\bauthor{\bsnm{{Hennebelle}}, \binits{P.}},
\bauthor{\bsnm{{Andr{\'e}}}, \binits{P.}}:
\bjtitle{\aap}
\bvolume{464},
\bfpage{983}
(\byear{2007}).
\arxivurl{astro-ph/0611277}.
doi:\doiurl{10.1051/0004-6361:20065653}
\end{barticle}
\endbibitem

\bibitem[\protect\citeauthoryear{{Rawlings} et~al.}{1992}]{raw92}
\begin{barticle}
\bauthor{\bsnm{{Rawlings}}, \binits{J.M.C.}},
\bauthor{\bsnm{{Hartquist}}, \binits{T.W.}},
\bauthor{\bsnm{{Menten}}, \binits{K.M.}},
\bauthor{\bsnm{{Williams}}, \binits{D.A.}}:
\bjtitle{\mnras}
\bvolume{255},
\bfpage{471}
(\byear{1992})
\end{barticle}
\endbibitem

\bibitem[\protect\citeauthoryear{{Roberts} and {Millar}}{2000a}]{rob20b}
\begin{barticle}
\bauthor{\bsnm{{Roberts}}, \binits{H.}},
\bauthor{\bsnm{{Millar}}, \binits{T.J.}}:
\bjtitle{\aap}
\bvolume{364},
\bfpage{780}
(\byear{2000}a)
\end{barticle}
\endbibitem

\bibitem[\protect\citeauthoryear{{Roberts} and {Millar}}{2000b}]{rob20a}
\begin{barticle}
\bauthor{\bsnm{{Roberts}}, \binits{H.}},
\bauthor{\bsnm{{Millar}}, \binits{T.J.}}:
\bjtitle{\aap}
\bvolume{361},
\bfpage{388}
(\byear{2000}b)
\end{barticle}
\endbibitem

\bibitem[\protect\citeauthoryear{{Roberts} et~al.}{2003}]{rob03}
\begin{barticle}
\bauthor{\bsnm{{Roberts}}, \binits{H.}},
\bauthor{\bsnm{{Herbst}}, \binits{E.}},
\bauthor{\bsnm{{Millar}}, \binits{T.J.}}:
\bjtitle{\apjl}
\bvolume{591},
\bfpage{41}
(\byear{2003}).
doi:\doiurl{10.1086/376962}
\end{barticle}
\endbibitem

\bibitem[\protect\citeauthoryear{{Roberts} et~al.}{2004}]{rob04}
\begin{barticle}
\bauthor{\bsnm{{Roberts}}, \binits{H.}},
\bauthor{\bsnm{{Herbst}}, \binits{E.}},
\bauthor{\bsnm{{Millar}}, \binits{T.J.}}:
\bjtitle{\aap}
\bvolume{424},
\bfpage{905}
(\byear{2004}).
doi:\doiurl{10.1051/0004-6361:20040441}
\end{barticle}
\endbibitem

\bibitem[\protect\citeauthoryear{{Roberts} et~al.}{2007}]{robj07}
\begin{barticle}
\bauthor{\bsnm{{Roberts}}, \binits{J.F.}},
\bauthor{\bsnm{{Rawlings}}, \binits{J.M.C.}},
\bauthor{\bsnm{{Viti}}, \binits{S.}},
\bauthor{\bsnm{{Williams}}, \binits{D.A.}}:
\bjtitle{\mnras}
\bvolume{382},
\bfpage{733}
(\byear{2007}).
\arxivurl{0708.3374}.
doi:\doiurl{10.1111/j.1365-2966.2007.12402.x}
\end{barticle}
\endbibitem

\bibitem[\protect\citeauthoryear{{Rodgers} and {Millar}}{1996}]{rod96}
\begin{barticle}
\bauthor{\bsnm{{Rodgers}}, \binits{S.D.}},
\bauthor{\bsnm{{Millar}}, \binits{T.J.}}:
\bjtitle{\mnras}
\bvolume{280},
\bfpage{1046}
(\byear{1996})
\end{barticle}
\endbibitem

\bibitem[\protect\citeauthoryear{{Sakai} et~al.}{2007}]{sak07}
\begin{barticle}
\bauthor{\bsnm{{Sakai}}, \binits{N.}},
\bauthor{\bsnm{{Sakai}}, \binits{T.}},
\bauthor{\bsnm{{Yamamoto}}, \binits{S.}}:
\bjtitle{\apj}
\bvolume{660},
\bfpage{363}
(\byear{2007}).
doi:\doiurl{10.1086/512774}
\end{barticle}
\endbibitem

\bibitem[\protect\citeauthoryear{{Sakai} et~al.}{2009}]{sak09}
\begin{barticle}
\bauthor{\bsnm{{Sakai}}, \binits{N.}},
\bauthor{\bsnm{{Sakai}}, \binits{T.}},
\bauthor{\bsnm{{Hirota}}, \binits{T.}},
\bauthor{\bsnm{{Yamamoto}}, \binits{S.}}:
\bjtitle{\apj}
\bvolume{702},
\bfpage{1025}
(\byear{2009}).
doi:\doiurl{10.1088/0004-637X/702/2/1025}
\end{barticle}
\endbibitem

\bibitem[\protect\citeauthoryear{{Saruwatari} et~al.}{2011}]{sar11}
\begin{barticle}
\bauthor{\bsnm{{Saruwatari}}, \binits{O.}},
\bauthor{\bsnm{{Sakai}}, \binits{N.}},
\bauthor{\bsnm{{Liu}}, \binits{S.-Y.}},
\bauthor{\bsnm{{Su}}, \binits{Y.-N.}},
\bauthor{\bsnm{{Sakai}}, \binits{T.}},
\bauthor{\bsnm{{Yamamoto}}, \binits{S.}}:
\bjtitle{\apj}
\bvolume{729},
\bfpage{147}
(\byear{2011}).
doi:\doiurl{10.1088/0004-637X/729/2/147}
\end{barticle}
\endbibitem

\bibitem[\protect\citeauthoryear{{Sipil{\"a}} et~al.}{2010}]{sip10}
\begin{barticle}
\bauthor{\bsnm{{Sipil{\"a}}}, \binits{O.}},
\bauthor{\bsnm{{Hugo}}, \binits{E.}},
\bauthor{\bsnm{{Harju}}, \binits{J.}},
\bauthor{\bsnm{{Asvany}}, \binits{O.}},
\bauthor{\bsnm{{Juvela}}, \binits{M.}},
\bauthor{\bsnm{{Schlemmer}}, \binits{S.}}:
\bjtitle{\aap}
\bvolume{509},
\bfpage{98}
(\byear{2010}).
\arxivurl{0911.1236}.
doi:\doiurl{10.1051/0004-6361/200913350}
\end{barticle}
\endbibitem

\bibitem[\protect\citeauthoryear{{Snow} and {McCall}}{2006}]{snow06}
\begin{barticle}
\bauthor{\bsnm{{Snow}}, \binits{T.P.}},
\bauthor{\bsnm{{McCall}}, \binits{B.J.}}:
\bjtitle{\araa}
\bvolume{44},
\bfpage{367}
(\byear{2006}).
doi:\doiurl{10.1146/annurev.astro.43.072103.150624}
\end{barticle}
\endbibitem

\bibitem[\protect\citeauthoryear{{Tielens}}{1983}]{tiel83}
\begin{barticle}
\bauthor{\bsnm{{Tielens}}, \binits{A.G.G.M.}}:
\bjtitle{\aap}
\bvolume{119},
\bfpage{177}
(\byear{1983})
\end{barticle}
\endbibitem

\bibitem[\protect\citeauthoryear{{Tielens}}{2013}]{tiel13}
\begin{barticle}
\bauthor{\bsnm{{Tielens}}, \binits{A.G.G.M.}}:
\bjtitle{Reviews of Modern Physics}
\bvolume{85},
\bfpage{1021}
(\byear{2013}).
doi:\doiurl{10.1103/RevModPhys.85.1021}
\end{barticle}
\endbibitem

\bibitem[\protect\citeauthoryear{{Tielens} and {Hagen}}{1982}]{tiel82}
\begin{barticle}
\bauthor{\bsnm{{Tielens}}, \binits{A.G.G.M.}},
\bauthor{\bsnm{{Hagen}}, \binits{W.}}:
\bjtitle{\aap}
\bvolume{114},
\bfpage{245}
(\byear{1982})
\end{barticle}
\endbibitem

\bibitem[\protect\citeauthoryear{{Tin{\'e}} et~al.}{2000}]{tin20}
\begin{barticle}
\bauthor{\bsnm{{Tin{\'e}}}, \binits{S.}},
\bauthor{\bsnm{{Roueff}}, \binits{E.}},
\bauthor{\bsnm{{Falgarone}}, \binits{E.}},
\bauthor{\bsnm{{Gerin}}, \binits{M.}},
\bauthor{\bsnm{{Pineau des For{\^e}ts}}, \binits{G.}}:
\bjtitle{\aap}
\bvolume{356},
\bfpage{1039}
(\byear{2000})
\end{barticle}
\endbibitem

\bibitem[\protect\citeauthoryear{{van der Tak} and {van
  Dishoeck}}{2000}]{vant20}
\begin{barticle}
\bauthor{\bsnm{{van der Tak}}, \binits{F.F.S.}},
\bauthor{\bsnm{{van Dishoeck}}, \binits{E.F.}}:
\bjtitle{\aap}
\bvolume{358},
\bfpage{79}
(\byear{2000}).
\arxivurl{astro-ph/0006246}
\end{barticle}
\endbibitem

\bibitem[\protect\citeauthoryear{{van der Tak} et~al.}{2009}]{vant09}
\begin{barticle}
\bauthor{\bsnm{{van der Tak}}, \binits{F.F.S.}},
\bauthor{\bsnm{{M{\"u}ller}}, \binits{H.S.P.}},
\bauthor{\bsnm{{Harding}}, \binits{M.E.}},
\bauthor{\bsnm{{Gauss}}, \binits{J.}}:
\bjtitle{\aap}
\bvolume{507},
\bfpage{347}
(\byear{2009}).
\arxivurl{0909.0390}.
doi:\doiurl{10.1051/0004-6361/200912912}
\end{barticle}
\endbibitem

\bibitem[\protect\citeauthoryear{{van Dishoeck} et~al.}{2003}]{vand03}
\begin{barticle}
\bauthor{\bsnm{{van Dishoeck}}, \binits{E.F.}},
\bauthor{\bsnm{{Thi}}, \binits{W.-F.}},
\bauthor{\bsnm{{van Zadelhoff}}, \binits{G.-J.}}:
\bjtitle{\aap}
\bvolume{400},
\bfpage{1}
(\byear{2003}).
\arxivurl{arXiv:astro-ph/0301571}.
doi:\doiurl{10.1051/0004-6361:20030091}
\end{barticle}
\endbibitem

\bibitem[\protect\citeauthoryear{{van Dishoeck} et~al.}{1995}]{vand95}
\begin{barticle}
\bauthor{\bsnm{{van Dishoeck}}, \binits{E.F.}},
\bauthor{\bsnm{{Blake}}, \binits{G.A.}},
\bauthor{\bsnm{{Jansen}}, \binits{D.J.}},
\bauthor{\bsnm{{Groesbeck}}, \binits{T.D.}}:
\bjtitle{\apj}
\bvolume{447},
\bfpage{760}
(\byear{1995}).
doi:\doiurl{10.1086/175915}
\end{barticle}
\endbibitem

\bibitem[\protect\citeauthoryear{{Vastel} et~al.}{2017}]{vas17}
\begin{barticle}
\bauthor{\bsnm{{Vastel}}, \binits{C.}},
\bauthor{\bsnm{{Mookerjea}}, \binits{B.}},
\bauthor{\bsnm{{Pety}}, \binits{J.}},
\bauthor{\bsnm{{Gerin}}, \binits{M.}}:
\bjtitle{\aap}
\bvolume{597},
\bfpage{45}
(\byear{2017}).
\arxivurl{1611.07389}.
doi:\doiurl{10.1051/0004-6361/201629289}
\end{barticle}
\endbibitem

\bibitem[\protect\citeauthoryear{{Viti} et~al.}{2004}]{viti04}
\begin{barticle}
\bauthor{\bsnm{{Viti}}, \binits{S.}},
\bauthor{\bsnm{{Collings}}, \binits{M.P.}},
\bauthor{\bsnm{{Dever}}, \binits{J.W.}},
\bauthor{\bsnm{{McCoustra}}, \binits{M.R.S.}},
\bauthor{\bsnm{{Williams}}, \binits{D.A.}}:
\bjtitle{\mnras}
\bvolume{354},
\bfpage{1141}
(\byear{2004}).
\arxivurl{arXiv:astro-ph/0406054}.
doi:\doiurl{10.1111/j.1365-2966.2004.08273.x}
\end{barticle}
\endbibitem

\bibitem[\protect\citeauthoryear{{Wakelam} et~al.}{2015}]{wak15}
\begin{barticle}
\bauthor{\bsnm{{Wakelam}}, \binits{V.}},
\bauthor{\bsnm{{Loison}}, \binits{J.-C.}},
\bauthor{\bsnm{{Herbst}}, \binits{E.}},
\bauthor{\bsnm{{Pavone}}, \binits{B.}},
\bauthor{\bsnm{{Bergeat}}, \binits{A.}},
\bauthor{\bsnm{{B{\'e}roff}}, \binits{K.}},
\bauthor{\bsnm{{Chabot}}, \binits{M.}},
\bauthor{\bsnm{{Faure}}, \binits{A.}},
\bauthor{\bsnm{{Galli}}, \binits{D.}},
\bauthor{\bsnm{{Geppert}}, \binits{W.D.}},
\bauthor{\bsnm{{Gerlich}}, \binits{D.}},
\bauthor{\bsnm{{Gratier}}, \binits{P.}},
\bauthor{\bsnm{{Harada}}, \binits{N.}},
\bauthor{\bsnm{{Hickson}}, \binits{K.M.}},
\bauthor{\bsnm{{Honvault}}, \binits{P.}},
\bauthor{\bsnm{{Klippenstein}}, \binits{S.J.}},
\bauthor{\bsnm{{Le Picard}}, \binits{S.D.}},
\bauthor{\bsnm{{Nyman}}, \binits{G.}},
\bauthor{\bsnm{{Ruaud}}, \binits{M.}},
\bauthor{\bsnm{{Schlemmer}}, \binits{S.}},
\bauthor{\bsnm{{Sims}}, \binits{I.R.}},
\bauthor{\bsnm{{Talbi}}, \binits{D.}},
\bauthor{\bsnm{{Tennyson}}, \binits{J.}},
\bauthor{\bsnm{{Wester}}, \binits{R.}}:
\bjtitle{\apjs}
\bvolume{217},
\bfpage{20}
(\byear{2015}).
\arxivurl{1503.01594}.
doi:\doiurl{10.1088/0067-0049/217/2/20}
\end{barticle}
\endbibitem

\bibitem[\protect\citeauthoryear{{Walmsley} et~al.}{2004}]{wal04}
\begin{barticle}
\bauthor{\bsnm{{Walmsley}}, \binits{C.M.}},
\bauthor{\bsnm{{Flower}}, \binits{D.R.}},
\bauthor{\bsnm{{Pineau des For{\^e}ts}}, \binits{G.}}:
\bjtitle{\aap}
\bvolume{418},
\bfpage{1035}
(\byear{2004}).
\arxivurl{arXiv:astro-ph/0402493}.
doi:\doiurl{10.1051/0004-6361:20035718}
\end{barticle}
\endbibitem

\bibitem[\protect\citeauthoryear{{Ward-Thompson} et~al.}{2000}]{war20}
\begin{barticle}
\bauthor{\bsnm{{Ward-Thompson}}, \binits{D.}},
\bauthor{\bsnm{{Zylka}}, \binits{R.}},
\bauthor{\bsnm{{Mezger}}, \binits{P.G.}},
\bauthor{\bsnm{{Sievers}}, \binits{A.W.}}:
\bjtitle{\aap}
\bvolume{355},
\bfpage{1122}
(\byear{2000})
\end{barticle}
\endbibitem

\bibitem[\protect\citeauthoryear{{Watanabe} and {Kouchi}}{2008}]{wat08}
\begin{barticle}
\bauthor{\bsnm{{Watanabe}}, \binits{N.}},
\bauthor{\bsnm{{Kouchi}}, \binits{A.}}:
\bjtitle{Progress In Surface Science}
\bvolume{83},
\bfpage{439}
(\byear{2008})
\end{barticle}
\endbibitem

\bibitem[\protect\citeauthoryear{{Watanabe} et~al.}{2003}]{wat03}
\begin{barticle}
\bauthor{\bsnm{{Watanabe}}, \binits{N.}},
\bauthor{\bsnm{{Shiraki}}, \binits{T.}},
\bauthor{\bsnm{{Kouchi}}, \binits{A.}}:
\bjtitle{\apjl}
\bvolume{588},
\bfpage{121}
(\byear{2003}).
doi:\doiurl{10.1086/375634}
\end{barticle}
\endbibitem

\bibitem[\protect\citeauthoryear{{Watanabe} et~al.}{2015}]{wat15}
\begin{barticle}
\bauthor{\bsnm{{Watanabe}}, \binits{Y.}},
\bauthor{\bsnm{{Sakai}}, \binits{N.}},
\bauthor{\bsnm{{L{\'o}pez-Sepulcre}}, \binits{A.}},
\bauthor{\bsnm{{Furuya}}, \binits{R.}},
\bauthor{\bsnm{{Sakai}}, \binits{T.}},
\bauthor{\bsnm{{Hirota}}, \binits{T.}},
\bauthor{\bsnm{{Liu}}, \binits{S.-Y.}},
\bauthor{\bsnm{{Su}}, \binits{Y.-N.}},
\bauthor{\bsnm{{Yamamoto}}, \binits{S.}}:
\bjtitle{\apj}
\bvolume{809},
\bfpage{162}
(\byear{2015}).
\arxivurl{1507.04958}.
doi:\doiurl{10.1088/0004-637X/809/2/162}
\end{barticle}
\endbibitem

\bibitem[\protect\citeauthoryear{{Watson}}{1980}]{wats80}
\begin{bchapter}
\bauthor{\bsnm{{Watson}}, \binits{W.D.}}:
In: \bbtitle{Les Spectres des Mol{\'e}cules Simples au Laboratoire et en
  Astrophysique},
p. \bfpage{526}
(\byear{1980})
\end{bchapter}
\endbibitem

\end{thebibliography}

\begin{table*}
   \centering
\caption {Initial physical conditions and chemical abundances utilized in this study for the RM of CMM3 in both Phase I, the dark cloud conditions, and Phase II, the protostellar core conditions.}
  \label{tab:initial}
    \leavevmode
    \begin{tabular}{lll} \hline \hline
\multicolumn{3}{c}{\bf Physical Conditions} \\ \hline
{\bf Parameters} & {\bf Phase I} & {\bf Phase II$^{\text a}$} \\ \hline
Core density (cm$^{-3}$) & 400 - 5.4$\times$10$^{6}$ & 5.4$\times$10$^{6}$ \\[0.8 ex]
Core temperature (K) & 10 & 15 \\[0.8 ex]
Core radius (pc) & 0.04 &  0.03 \\[0.8 ex]
$^{\dag}$Depletion & full & -- \\[0.8 ex]
$^{\dag\dag}\zeta$ (s$^{-1}$) & 1.3 $\times$ 10$^{-17}$ &  1.69 $\times$ 10$^{-17}$ \\[0.8 ex] \hline \hline 
\multicolumn{3}{c}{\bf Chemical Abundances} \\ \hline
{\bf Species}&{\bf Phase I$^{\text b}$} & {\bf $^\ddag$Phase II} \\\hline
Helium & 8.50 $\times$ 10$^{-2}$ & --- \\[0.8 ex]
Carbon & 2.69 $\times$ 10$^{-4}$ & --- \\[0.8 ex]
Oxygen & 4.90 $\times$ 10$^{-4}$ & --- \\[0.8 ex]
Nitrogen & 6.76 $\times$ 10$^{-5}$& --- \\[0.8 ex]
$\ddag\ddag$HD & 1.5 $\times$ 10$^{-5}$& ---  \\[0.8 ex]
CO & ---  & 5.4 $\times$ 10$^{-12}$ \\[0.8 ex]
N$_2$ & --- & 4.2 $\times$ 10$^{-10}$ \\[0.8 ex]
H$_2$D$^+$ & --- & 6.6 $\times$ 10$^{-12}$ \\[0.8 ex]
C$_2$HD$^+$ & --- & $<$ 1.0 $\times$ 10$^{-13}$ \\[0.8 ex]
CH$_2$D$^+$ & --- & $<$ 1.0 $\times$ 10$^{-13}$  \\ \hline \hline
\end{tabular}
\flushleft
$\dag$ This parameter is related to the freeze-out process occur in the core, only in the pre-phase. The gas is considered fully depleted if more than 90\% of the species gas-phase abundances are involved into grain surface reactions.\\
$\dag\dag$ Dark cloud chemistry runs under the standard CR ionisation rate, $\zeta_{\text{ISM}}$ = 1.3 $\times$ 10$^{-17}$ s$^{-1}$, while in the protostellar phase, this value is enhanced by 1.3 times \citep{awad17}.\\
$\ddag$ The low values of the initial molecular abundances in Phase II is due to their depletion onto grain surfaces (see \S\ref{mod} and \S\ref{res}). \\
$\ddag\ddag$ HD molecules represent the abundance of atomic D in the gas following \citet{awad14}; given that D/H ratio = 10$^{-5}$ \citep{oli03}. \\
References: (a) \citealt{per06}, (b) \citealt{asp09}
\end{table*}
\begin{table*}
   \centering
\caption{The grid of models studied in the present work. The table lists the main differences in the physical parameters compared to the reference model (RM), keeping the rest of the parameters stated in Table \ref{tab:initial} unchanged.}
  \label{tab:grid}
    \leavevmode
    \begin{tabular}{llll} \hline \hline
{\bf Model} & {\bf Density} & {\bf $^{\dag}$Depletion} & {\bf $^{\ddag}\zeta_{\text{CMM3}}$}\\[0.8 ex]
& (cm$^{-3}$) &  & ($\zeta_{\text{ISM}}$ s$^{-1}$)\\[0.8 ex] \hline
RM & 5.4$\times$10$^{6}$ & full & 1.3 \\[0.8 ex]
M1 & 1.0$\times$10$^{6}$ & full & 1.3 \\[0.8 ex]
M2 & 2.7$\times$10$^{7}$ & full & 1.3 \\ [0.8 ex]
M3 & 5.4$\times$10$^{6}$ & partial & 1.3 \\[0.8 ex]
M4 & 5.4$\times$10$^{6}$ & none & 1.3 \\[0.8 ex]
M5 & 5.4$\times$10$^{6}$ & full & 5 \\[0.8 ex]
M6 & 5.4$\times$10$^{6}$ & full & 10 \\[0.8 ex] \hline \hline 
\end{tabular}
\flushleft
$\dag$ Partially depleted gas means, in this study, that $\sim$ 85\% of the amount of the gaseous species are involved in grain surface reactions, while none-depleted gas represents the pure gas-phase chemistry in which 0\% of the gas is involved in surface reactions.\\
$\ddag$ The cosmic ray ionisation rate in CMM3 in units of the standard rate; $\zeta_{\text{ISM}}$. 
\end{table*}
\begin{table*}
\centering
\caption{Comparison between our calculated abundances and D/H ratios from the reference model (RM) and observations of selected deuterated species in NGC 2264 CMM3 from Tables 6 and 7 in \citet{wat15} at 15 K. Time of best fit, in years, is also shown in the last column.}
  \label{tab:comp}
    \leavevmode
    \begin{tabular}{llllllll} \hline\hline
\multicolumn{8}{c}{\bf \underline {Fractional Abundances}} \\ [0.8 ex]
{\bf Species} & \multicolumn{3}{c}{\underline{\bf Observations$^{\dag}$}} && {\bf This Work} && \multicolumn{1}{l}{\bf Time} \\ [0.8 ex] 
              &  \multicolumn{1}{c}{\bf N(Y)} &&   \multicolumn{1}{c}{\bf $x$(Y)} &&    \multicolumn{1}{c}{\bf $x$(Y)} && \multicolumn{1}{l}{\bf in yrs} \\ \hline
{\bf DCO$^+$} & 7.4 $\times$ 10$^{13}$  && 6.5 $\times$ 10$^{-11}$ && 6.2 $\times$ 10$^{-11}$ && $\ge$ 6.4 $\times$ 10$^{4}$ \\ [0.8 ex]  
{\bf N$_2$D$^+$} & 7.8 $\times$ 10$^{12}$  && 6.8 $\times$ 10$^{-12}$ && $<$ 1.0$\times$ 10$^{-12}$ && all times \\ [0.8 ex]  
{\bf DCN} & 3.6 $\times$ 10$^{13}$  && 3.2 $\times$ 10$^{-11}$ && 2.7 $\times$ 10$^{-11}$ && 2.4 $\times$ 10$^{4}$ \\ [0.8 ex]  
{\bf DNC} & 2.4 $\times$ 10$^{13}$  && 2.1 $\times$ 10$^{-11}$ && $\sim$ 1.0 $\times$ 10$^{-11}$ && $>$ 3.8 $\times$ 10$^{5}$ \\ [0.8 ex]  
{\bf C$_2$D} & 1.3 $\times$ 10$^{14}$ && 1.1 $\times$ 10$^{-10}$ && $\ge$ 1.0 $\times$ 10$^{-12}$ && all times \\ [0.8 ex] 
{\bf HDCO} & 1.0 $\times$ 10$^{13}$ && 8.8 $\times$ 10$^{-12}$ && $\ge$ 4.0 $\times$ 10$^{-13}$ && all times \\ \hline \hline
\multicolumn{8}{c}{\bf \underline {D/H ratios}} \\ [0.8 ex]
\multicolumn{1}{l}{\bf The ratio} &&\multicolumn{2}{l}{\bf Observations} && \multicolumn{1}{l}{\bf This Work}&& \multicolumn{1}{l}{\bf Time (in yrs)} \\  \hline
{\bf DCO$^+$/HCO$^+$}        && 0.030 &$\pm$ 0.015 && 0.005 - 0.05 && (1 - 4) $\times$ 10$^4$\\[0.8 ex]
{\bf N$_2$D$^+$ /N$_2$H$^+$} && 0.026 &$\pm$ 0.010 && $\le$ 0.05  && $\ge$ 1 $\times$ 10$^4$\\[0.8 ex]
{\bf DCN/HCN}               && 0.010 &$\pm$ 0.003 && 0.02 - 0.03 && (1 - 4.5) $\times$ 10$^4$ \\[0.8 ex]
{\bf DNC/HNC}               && 0.017 &$\pm$ 0.012 && $>$ 0.05 && all times \\[0.8 ex]
{\bf C$_2$D/C$_2$H}         && 0.037 &$\pm$ 0.032 && $\ge$ 0.2 && all times \\[0.8 ex]
{\bf HDCO/H$_2$CO}        && 0.0020 &$\pm$ 0.0015 && $\ge$ 0.1 && all times \\[0.8 ex]
\hline\hline 
\end{tabular}
\flushleft
$^{\dag}$ Observed column densities, N(Y), are converted into fractional abundances with respect to H, $x$(Y), given the observed N(H$_2$) in NGC 2264 CMM3 is 5.7$\times$10$^{23}$ cm$^{-2}$ \citep{per06}.\\
\end{table*}

\end{document}